\documentstyle[prl,aps,twocolumn,psfig,floats]{revtex}

\begin{document}

\title{An investigation of standard thermodynamic quantities as determined via models of nuclear multifragmentation}

\author{
J. B. Elliott$^1$ and A. S. Hirsch$^2$
}

\address{
$^1$Nuclear Science Division, Lawrence Berkeley National Laboratory, Berkeley, California 94720\\
$^2$Purdue University, West Lafayette, IN 47907
}

\date{\today}

\maketitle

\begin{abstract}
Both simple and sophisticated models are frequently used in an attempt to understand how real nuclei breakup when subjected to 
large excitation energies, a process known as nuclear multifragmentation.\ \ Many of these models assume equilibrium 
thermodynamics and produce results often interpreted as evidence of a phase transition.\ \ This work examines one class of 
models and employs standard thermodynamical procedure to explore the possible existence and nature of a phase transition.\ \ 
The role of various terms, {\it e.g.} Coulomb and surface energy, is discussed.
\end{abstract}
\narrowtext

\section{Introduction and Overview}

Many models \cite{bondorf} - \cite{botvina} have been proposed the describe the breakup of a large nucleus subjected to excitation 
energies greater than a few MeV per nucleon, a process known as multifragmentation.\ \ Experimentally, the signature of 
multifragmentation is the production of a wide range of nuclear reaction products, particularly intermediate mass fragments 
(IMFs), $ 3 \le Z \le 30$.\ \ On the basis of inclusive data, it was proposed \cite{finn} - \cite{hirsch} that these fragments 
were produced in analogy to a liquid-to-gas phase transition occurring in a nucleus.\ \ A recent experiment that permitted the 
total charge reconstruction of each event studied multifragmentation resulting from the breakup of gold nuclei as a function of 
the excitation energy deposited \cite{gilkes_gamma} - \cite{elliott_scaling}.\ \ The statistical aspects of these data have 
provided strong evidence that multifragmentation is indeed related to a phase transition occurring in a finite system.\ \ Whether 
the production of IMFs in such collisions is due to a phase transition, and if so, what type, is still an issue of much debate 
\cite{gsi_caloric}.

One class of models developed to explore the fate of a nucleus as a function of excitation energy is based on the phenomenological 
description of the free energy, $F(V,T)$, of the breakup state, where $T$ is the common temperature of all nucleons and nuclei within 
the breakup volume $V$.\ \ These nuclei are considered to be at normal nuclear density and interact only via the Coulomb force.\ \ The 
distribution of nuclear fragments prior to any secondary decay can then be calculated as a first step in the disassembly of the excited 
initial system.\ \ To compare with data, deexcitation of fragments and expansion of the system due to the Coulomb repulsion between the 
fragments must be accounted for in the model.\ \ However, if the thermodynamics of the model is of interest, as is the case in this work, 
then only the behavior of thermodynamic variables need be examined, {\it e.g.} free energy, entropy, specific heat, pressure, isothermal 
compressibility.\ \ Thus, no fragment distributions need be explicitly calculated and therefore no fragment distributions are analyzed in 
this work.

Here, several variations of a previously discussed model \cite{mdgm} are explored.\ \ A canonical ensemble approach is used to 
investigate the thermodynamics of the system where the free energy, $F$, is written as a function of the temperature, $T$, and 
the volume, $V$.\ \ Calculations are restricted to a system which contains 162 constituent, since this is representative of the 
size of the system studied in \cite{hauger_prc}.\ \ Contributions to $F$, {\it e.g.} the surface free energy, the Coulomb energy, 
are examined by turning them off or altering the form of the contribution in question.\ \ In this way insight can be gained as to 
how the important features of the thermodynamics such as specific heat, isothermal compressibility, etc. depend on the parameterization 
of the free energy.

This paper is organized as follows.\ \ In section II the details of the models are presented.\ \ Three versions of a standard
statistical multifragmentation model are examined as well as a well-known mean field model whose results are used for comparison.\ \ 
In section III a description of the analysis and the results of that analysis are presented.\ \ Section IV discusses the standard 
interpretations of the models and analysis.\ \ Finally, a brief discussion of the  questions raised by this work concludes this 
paper.\ \ In general, the notation of references \cite{bondorf} and \cite{mdgm} are followed.

\section{Details of the models}

This work follows directly the efforts presented in reference \cite{mdgm} in which the canonical partition function was examined as
a function of temperature in a fixed volume system for evidence of a phase transition.\ \ In that work, evidence for a first order
phase transition was found.\ \ In the present work, the volume (average density) of the system is permitted to vary.\ \ It shall be seen 
that the nature of the phase transition depends on the volume of the system.\ \  The work of \cite{mdgm} is also extended by 
examining the effects of the Coulomb force on the system and the effects of the choice of surface energy parameterizations.\ \ The 
units used for the nuclear models are: energy and free energy in MeV/nucleon, temperature in MeV, volume in fm$^3$, pressure in 
MeV$/$fm$^3$ and so on.

A general description of each system follows.

\subsection{V1: Full statistical description of an excited nucleus}

Calculations begin by considering the free energy of a nuclear fragment.\ \ It is assumed \cite{bondorf} that the free energy of a
nuclear fragment of mass $A$ and charge $Z$, for $A > 1$ is given by:
	\begin{equation}
		F_{A,Z} = F_{A,Z}^{B} + F_{A,Z}^{sym} + F_{A,Z}^{S} + E_{A,Z}^{C}.
	\label{F_AZ}
	\end{equation}
The terms in eq. (\ref{F_AZ}) refer to the bulk, symmetry, surface and Coulomb contributions to the free energy of a nuclear
fragment.\ \ The forms of these terms are given \cite{bondorf}:
	\begin{equation}
	 	F_{A,Z}^{B} = ( -W_{0} - T^{2} / {\epsilon}_{0} ) A ,
	\label{F_AZ_B}
	\end{equation}
	\begin{equation}
		F_{A,Z}^{sym} = {\gamma} ( A - 2 Z )^{2} / A ,
	\label{F_AZ_sym}
	\end{equation}
	\begin{equation}
		F_{A,Z}^{S} = {\beta}_{0} \left( \frac{T_{c}^{2}-T^{2}}{T_{c}^{2}+T^{2}} \right)^{5/4} A^{2/3},
	\label{F_AZ_S}
	\end{equation}
	\begin{equation}
		E_{A,Z}^{C} = \frac{3}{5} e^{2} Z^{2} (1-(1+{\kappa})^{-1/3})  / R_{A,Z} .
	\label{E_AZ_C}
	\end{equation}
In eq. (\ref{F_AZ_B}) the constants are taken as $W_0 = 16$ MeV and ${\epsilon}_0 = 16$ MeV.\ \ In eq. (\ref{F_AZ_sym}) ${\gamma} =
25$ MeV.\ \ In eq. (\ref{F_AZ_S}) ${\beta}_0 = 18$ MeV and $T_c = 16$ MeV, following reference \cite{botvina}.\ \ The contribution
from the Coulomb term is estimated via a Wigner-Seitz approximation as in reference \cite{bondorf}.

The $\kappa$-term is related to the volume of the system through
	\begin{equation}
		1 + {\kappa} = V / V_0 .
	\label{kappa}
	\end{equation}
This simplified model presented here differs from the standard version \cite{bondorf} in that there is only one parameter relating 
the volume excluded by the constituents, $V_0$, to the total volume of the system, $V$, and to the free volume, $V_f$.\ \ Here the 
free volume is the difference between the total volume $V$ and the sum of the volume of the fragments, assumed to be at normal nuclear 
density, and is the volume available for the translational motion of the fragments.

In the standard version of the model the free volume is given by $V_f = {\chi} V_0$, where $\chi$ is parameterized to increase with 
fragment multiplicity such that it varies between $0.2$ and $2$; the parameter $\kappa$ is fixed, usually at ${\kappa} = 2$.\ \ For 
simplicity, here it is assumed that ${\kappa} = {\chi}$ so that specifying $V_f$ determines the value of $\kappa$ in eq. 
(\ref{kappa}).\ \ See reference \cite{bondorf} for details of $\kappa$ and $\chi$ in the standard version.

For this work then, the total volume of the system is then given by:
	\begin{equation}
		V = V_0 + V_f .
	\label{v_f}
	\end{equation}
Two things become obvious from eq. (\ref{v_f}); first, with this form of $V$ the free energy of the system varies with $V_f$ 
since $V_0$ is a constant.\ \  Second, the loss of free volume in the closest packing of spherical clusters is ignored.\ \ The 
issue of whether spherical nuclei can actually be placed in a total volume, $V_0$, given a free volume, $V_f$, is not addressed.\ \ 
Undoubtedly, there will be situations where it is not possible for the total volume to accommodate all of the nuclear clusters.\ \ 
The purpose here, however, is to explore the thermodynamics and self-consistency of the model and not physical consistency.

Finally $R_{A,Z}$ is the radius of the fragment in question and is determined by
	\begin{equation}
		R_{A,Z} = r_0 A^{1/3} ,
	\label{R_AZ}
	\end{equation}
with $r_0 = 1.17$ fm.\ \ The version of the model presented above will be termed V1.

From this point the intrinsic partition function of a fragment of $A$, $Z$ at temperature $T$ and volume $V$ can be determined
\cite{mdgm} as follows:
	\begin{equation}
		z_{A,Z} = \exp ( - F_{A,Z} / T ) .
	\label{frag_part_fcn}
	\end{equation}
Using a technique developed in reference \cite{mekjian} and used on a simplified version of this model \cite{mdgm}, the canonical
partition function can be built via a recursion relation:
	\begin{equation}
		{\cal Z}_{p} = \frac{1}{p} \sum_{A=1}^{p} A {\omega}_A {\cal Z}_{p-A} ,
	\label{part_fcn}
	\end{equation}
starting from ${\cal Z}_0 = 1$.\ \ Here for calculational simplicity the approximation has been made that for each and every
fragment with $A > 1$, $A / Z = 2.5$ which represents an average mass to charge ratio for fragments.\ \ The ${\omega}_A$ term is
	\begin{equation}
		{\omega}_{A} = \frac{V_f}{h^3} ( 2 {\pi} m T )^{3/2} A^{3/2} z_{A,Z} ,
	\label{omega}
	\end{equation}
where the terms the the left of the fragment partition function, $z_{A,Z}$, account for the translational free energy
contribution, $F^{tr}$.\ \ It is now straightforward to calculate the partition function of the system for a given $T$, $V_f$, $A_0$ and
$Z_0$.\ \ The free energy of the system of $p$-particles is then determined as usual
	\begin{equation}
		F = - T \log ( {\cal Z}_{p} ) + E_{0}^{c}(V) ,
	\label{F_tot}
	\end{equation}
where the last term is the usual Coulomb contribution of a uniformly charged sphere:
	\begin{equation}
		E_{0}^{c}(V) = \frac{3}{5} \frac{Z_{0}^{2} e^{2}}{R} ,
	\label{coulomb}
	\end{equation}
with $R = ( 3 V / 4 {\pi} )^{1/3}$ .

\subsubsection{Comparison of V1 to the full version of the model}

The model and calculations described above were compared to the full, or unmodified version of the model often cited in the literature, see 
for example references \cite{bondorf}, \cite{bondorf1} - \cite{agostino}.\ \  See Figure 1.\ \ In Figure 1 results from the full version 
of this model are shown for the mean fragment distribution calculated at a given input excitation energy.\ \ To generate event-by-event 
distributions Poissonian fluctuations about the mean are introduced, after which, temperature is adjusted to ensure energy conservation.\ \ 
To more fully recover the standard version of the model most often used, higher order corrections were introduced just as in the full 
version of the model; {\it e.g.} ${\epsilon}_0$ in eq. (\ref{F_AZ_B}) was made dependent on the fragment mass $A$, for light 
clusters, $A \le 4$, the empirical masses and binding energies, radii and spin degeneracy factors of the ground state were used, 
the total volume was held constant at $3 V_0$ and the free volume was set to depend on the input excitation energy.\ \  Finally,
energy was explicitly conserved; an input excitation energy was given and a temperature was determined such that total energy was 
conserved.\ \ The explicit conservation of energy produced results that were essentially the same as those resulting from the 
unconstrained canonical ensemble.

\begin{figure} [ht]
\centerline{\psfig{file=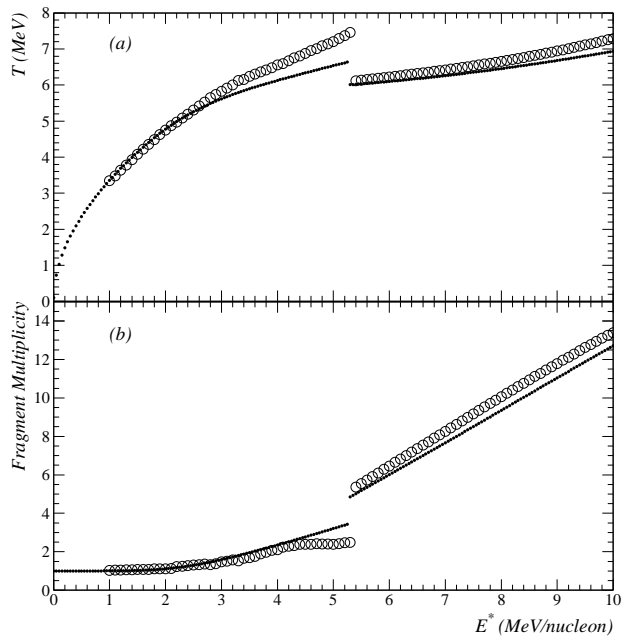,width=9.6cm,angle=0}}
\caption{Comparison of the full model (open circles) before Poissonian smearing and the calculations of this paper (full circles) 
for the caloric curve (a) and the average fragment multiplicities (b).}
\label{fig:00}
\end{figure}

Figure 1a shows the caloric curve from the full version of this model for a system with 100 nucleons (60 neutrons and 40 protons) 
compared to the same size system used in calculations with a modified version of V1.\ \ The general trend of the modified V1 
reproduced the average behavior of the full model, though there is not a perfect agreement.\ \ This is to be expected.\ \ While this
modified version of V1 is closer to the model, there are still some differences, {\it e.g.} the charge of $A > 4$ fragment is treated in 
only an average fashion in V1.\ \ The reproduction of the general trends indicates that V1 captures the essence of the full model.\ \ Figure 
1b shows the fragment multiplicity, before any secondary decay, from both models.\ \ Again there is general agreement between the two.

The break observed in the caloric curve shown in Figure 1 is well known in the full model, see, for example, Figure 4 in ref. 
\cite{bondorf1} and Figure 11 in ref. \cite{agostino}.\ \ The break is due to the initial guess of the system's multiplicity which 
is in turn used to guess the system's free volume.\ \ For low energies the multiplicity is chosen to be 1, 2 or 3 (Figure 1b shows
that the initial guess of the multiplicity is consistent with the final state multiplicity), while at higher energies the multiplicity 
depends smoothly on a function of the input excitation energy \cite{bondorf}.\ \ In some systems, {\it e.g.} a system of 100 nucleons, 
there is a jump in multiplicity at the transition from the low energy computations and the high energy computations which gives rise 
to a jump in the final state multiplicity, Figure 1b, and a break in the caloric curve, Figure 1a.\ \ When the simplified model used 
in this work is given the same volume dependence as the full model, the results of the full model are reproduced.

Energy conservation is explored in Figure 2.\ \ Here the unmodified version of V1 was used with a system of 162 particles and energy 
was explicitly conserved as outlined above.\ \ In order to examine the change in energy between the initial and final state of each 
term contributing to the system's total energy, the temperature of the initial state of the system has been calculated corresponding 
to the input $E^*$.\ \  The assumption that the initial nuclear state is in thermal equilibrium prior to its deexcitation to the final 
state has no bearing on the thermodynamics of the final state and is done only for purposes of the abovementioned calculation.\ \ The 
initial state used for this calculation was the system of 162 nucleons at excitation energy, $E^*$, at normal nuclear density, 
${\rho}_0$ and at a temperature that conserves energy when the total energy is determined using eqns (\ref{F_AZ_B})-(\ref{E_AZ_C}), 
(\ref{F_tot}) and:
	\begin{equation}
		E = F - T \left( \frac{\partial F}{\partial T} \right) _V .
	\label{E_total}
	\end{equation}
The total energy of the initial state is shown in Figure 2 as well as it's various components and a caloric curve.\ \ The 
final state of the system was computed with the same $E^*$ but was held at a third normal density, ${\rho}_0 / 3$, and allowed to 
fragment in the manner outlined above.\ \ The caloric curve produced for the final state via this explicit conservation of energy 
calculation is identical to the caloric curve produced via the calculations without an explicit conservation of energy.\ \ Figure 
2 shows that the temperature in the final state is lower than that in the initial state.\ \ Further inspection indicates that while 
the Coulomb energy is reduced by creating smaller charged nuclei, the energy required to create the additional surface area more is 
more than offsetting.\ \ Thus, the temperature must decrease.

\begin{figure} [ht]
\centerline{\psfig{file=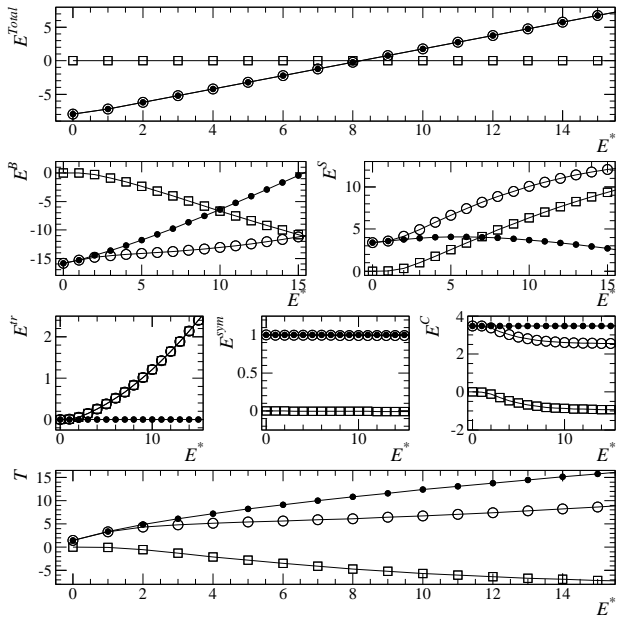,width=9.6cm,angle=0}}
\caption{Energy conservation in V1, overall, component by component (total, bulk, surface, translational, symmetry and Coulomb energies 
respectively) and the caloric curves.\ \ Full circles represent the initial state, open circles represent the final state and open 
squares represent the difference between between the two.}
\label{fig:0a}
\end{figure}

\subsection{V2: Description of an excited charge-free nucleus}

The general ideas of V1 are followed but with the Coulomb energy of the system set to zero.\ \ The free energy shown in 
eq. (\ref{F_AZ}) then becomes:
	\begin{equation}
		F_{A} = F_{A}^{B} + F_{A}^{S}.
	\label{F_A_V2}
	\end{equation}
And the total free energy of the system is given by 
	\begin{equation}
		F = - T \log ( {\cal Z}_{p} ) .
	\label{F_tot_V2}
	\end{equation}
Every other aspect of the model is the same as in V1.\ \ This version of the model will be termed V2.

\subsection{V3: Description of an excited charge-free nucleus with a temperature independent surface}

In this version, the Coulomb force is suppressed and the temperature dependent surface term in eq. (\ref{F_AZ_S}) is made
independent of temperature:
	\begin{equation}
		F_{A}^{S} = {\beta}_{0} A^{2/3}.
	\label{F_A_S_V3}
	\end{equation}
The free energy of a fragment and the entire system are still given by eqns (\ref{F_A_V2}) and (\ref{F_tot_V2}).\ \ Every other
aspect of the model is the same as in V1 and V2.\ \ This version of the model will be termed V3. 

\subsection{The van der Waals fluid}

The free energy of the van der Waals fluid is determined in the standard textbook fashion.\ \ Starting from the free energy
\cite{kittel}
	\begin{equation}
		F / A = - t \left\{ \ln \left[ n_{Q} \left( V - A b \right) / A \right] + 1 \right\} - A a / V,
	\label{FE_vdw}
	\end{equation}
with $t = k_{b} T $, $n_{Q} = \left( M t / 2 {\pi} {\hbar}^{2} \right) ^{3/2}$ and $a$ and $b$ the usual van der Waals constants.\
\ In this work, eq. (\ref{FE_vdw}) is computed in terms of the density, ${\rho} = A / V$, so that the number of constituents, A, is
not a factor.\ \ For the van der Waals' constants $a$ and $b$, values were used for helium so that $T_c \sim 4.5 \times 10^{-4}$ 
eV$/$A.\ \ This also suggested the value of $M$ in $n_Q$.\ \ Finally, eq. (\ref{FE_vdw}) was evaluated in terms of $T/T_c$ and $V/V_c$.\ \ 
The van der Waals fluid model is well defined ad free of internal inconsistencies.\ \ It will be used as a benchmark for the analysis 
presented in this paper.\ \ Units for the van der Waals fluid results will be in eV$/A$ for energy and free energy, eV$/A \times V_c$ 
for pressure and so on.

\section{Discussion of calculations}

Calculations were performed for each version of the model to determine $F(T , V_f , A_0 , Z_0 )$ for $A_0 = 162$, $A_0 / Z_0 = 2.5$
and over a range in temperature, $1$ MeV $ \le T \le 14$ MeV, and volume of $2 \times 10^{-8} \le ( V_{f} / V_{0} ) \le$ $2 \times
10^{8}$.\ \ Once the general vicinity of the critical point was identified, a smaller range in $(T,V_f )$ was used for more detailed 
calculations.\ \ For the van der Waals fluid the range was smaller and in terms of reduced temperature and volume:  
$ 0.1 \le T/T_c \le 2.0 $ and $0.34 \le V/V_c \le 2.0$.\ \ Figure 3 shows the behavior of the free energy over the ranges of 
temperature and volume used in the calculations.

\begin{figure} [ht]
\centerline{\psfig{file=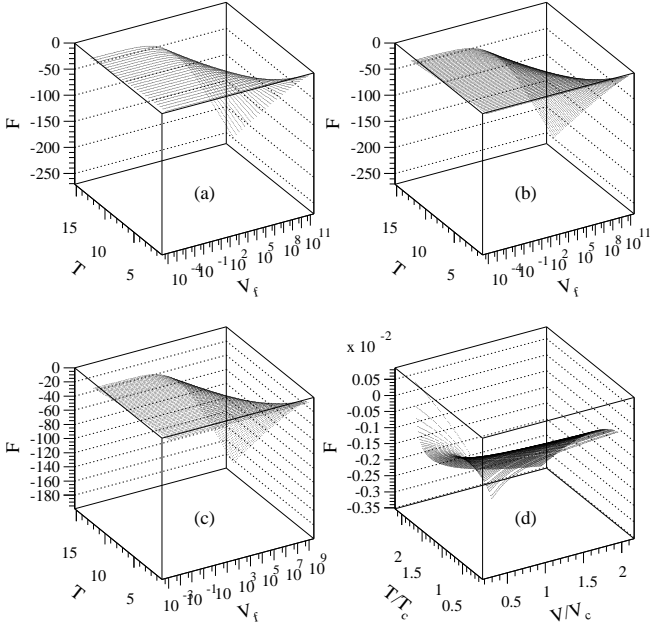,width=9.6cm,angle=0}}
\caption{Free energy surfaces as a function of temperature and volume for (a) V1, (b) V2, (c) V3 and (d) the van der Waals fluid.}
\label{fig:01}
\end{figure}

After the value of the free energy was calculated, it was simple to determine other thermodynamic quantities.\ \ Holding the 
volume fixed the entropy is given by the usual relation:
	\begin{equation}
		S = - \left( \frac{{\partial} F}{{\partial} T} \right)_{V_{f}} .
	\label{entropy1}
	\end{equation}
In the case of this work differences were used instead of derivatives due to the numerical nature of the calculation, thus eq.
(\ref{entropy1}) becomes
	\begin{equation}
		S = - \left( \frac{{\Delta} F}{{\Delta} T} \right)_{V_{f}} .
	\label{entropy2}
	\end{equation}
Similarly the specific heat at constant volume was determined via
	\begin{equation}
		C_V =  \left<T\right> \left( \frac{{\Delta} S}{{\Delta}T} \right)_{V} ,
	\label{cv}
	\end{equation}
where $\left<T\right>$ is the average value of $T$ over the ${\Delta} T$ interval.\ \ Using the entropy and the free energy, the total energy
can be determined from:
	\begin{equation}
		E = F + T S .
	\label{energy}
	\end{equation}

In these calculations it was possible to hold either the temperature or the volume constant.\ \ The pressure was then found by 
holding the temperature fixed
	\begin{equation}
		P = - \left( \frac{{\Delta} F}{{\Delta} V_f } \right)_{T} .
	\label{pressure}
	\end{equation}
Taking another {\it derivative} then gave the isothermal compressibility
	\begin{equation}
		{\kappa}_T = - \frac{1}{\left< V_f \right>} \left( \frac{{\Delta} V_f }{{\Delta} P} \right)_{T} .
	\label{iso_comp}
	\end{equation}
where $\left< V_f \right>$ is the average value of $V_f$ over the ${\Delta} V_f$ interval.\ \ With this information it is possible to
determine if there is a phase transition in a model such as this and, if present, the nature of that phase transition.\ \ The
following section addresses this question.

\section{Results of calculations}

In this section, each of the axes of the standard phase diagram, $T$, $V$ and $P$, will in turn be held fixed.\ \ The behavior of other
quantities will be examined in order to understand the behavior of each system.\ \ The van der Waals fluid will serve as a guide 
for the interpretation of the analysis and also as a benchmark to illustrate the accuracy of this analysis.

\subsection{Isotherms}

Determination of the critical point, coexistence and spinodal curves is discussed below.\ \ The variation of the free energy as a 
function of density shows the same general features for these systems.

\begin{figure} [ht]
\centerline{\psfig{file=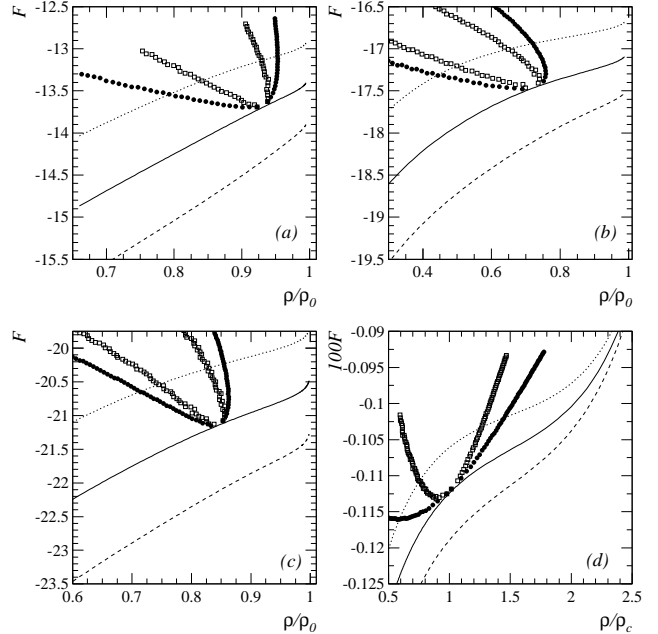,width=9.6cm,angle=0}}
\caption{Free energy isotherms as a function of reduced density for (a) V1, (b) V2, (c) V3 and (d) the van der Waals fluid.\ \
Open squares show the spinodal, full circles show the boundary of the coexistence region.\ \ Dotted curves are sub-critical.\ \
Solid curves are critical.\ \ Dashed curves are super-critical.}
\label{fig:02}
\end{figure}

Figure 4 shows the behavior of the free energy isotherms for each system as function of reduced density.\ \ Each plot in Figure 3
shows three isotherms.\ \ The isotherms are sub-critical ($T=0.95T_c$), critical ($T=T_c$) and super-critical ($T=1.05T_c$).\ \ 
Determination of the critical point is discussed below.\ \ Also shown are the approximate location of the coexistence and 
spinodal curves.\ \ Determination of these curves is also discussed below.\ \ The behavior of the free energy for these isotherms 
for all systems is more or less the same.\ \ As the reduced density increases, the free energy increases.\ \ At some mid range in 
reduced density the slope of the increase in free energy changes.\ \ At a greater reduced density the slope of the increase in 
free energy changes again.\ \ This is most clearly demonstrated by the van der Waals fluid system.\ \ See Figure 4d.\ \ However, 
the behavior is present in all the models.  This behavior, while appearing modest in these plots, will be seen to be the cause of 
the critical-like behavior exhibited by these models.

\begin{figure} [ht]
\centerline{\psfig{file=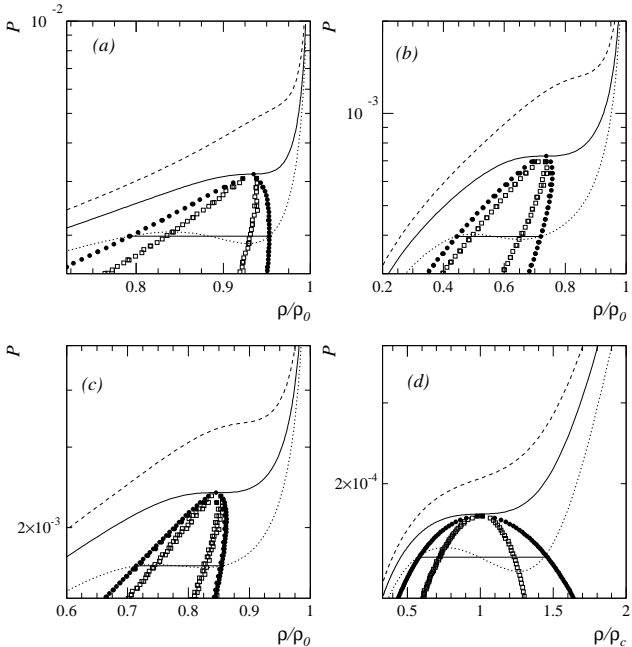,width=9.6cm,angle=0}}
\caption{Isotherms or pressure as a function of reduced density for (a) V1, (b) V2, (c) V3 and (d) the van der Waals fluid.\ \
Open squares show the spinodal, full circles show the coexistence curve.\ \ The solid horizontal line is an example of a Maxwell
equal area construction to determine the coexistence region and find the critical point.\ \ Dotted curves are sub-critical.\ \
Solid curves are critical.\ \ Dashed curves are super-critical.}
\label{fig:03}
\end{figure}

The pressure was calculated from the free energy isotherms via eq. (\ref{pressure}).\ \ Figure 5 shows the results for each 
model.\ \ The determination of the location of the critical point and the coexistence and spinodal curves is based on the 
phase diagram of pressure, temperature and reduced density.\ \ By searching for inflections points along the pressure 
versus reduced density isotherms the spinodal curve was determined.\ \ The isotherm immediately following, as the temperature of 
each isotherm increases, the last isotherm with two inflection points was labeled the critical isotherm.

Another method to determine the location of the critical point began with the isothermal compressibility which was calculated 
with eq. (\ref{iso_comp}).\ \ Isotherms of ${\kappa}_T$ versus reduced density were inspected.\ \ All sub-critical isotherms 
showed at least one negative value of ${\kappa}_T$.\ \ The first isotherm, as a function of increasing temperature, which showed only 
positive values of ${\kappa}_T$ was labeled the critical isotherm.\ \ Both procedures yielded the same results, as they are 
essentially identical.\ \ See Table I.

The coexistence curve was determined by making a Maxwell equal area construction for each isotherm.\ \ See Figure 4 for examples of 
the Maxwell construction.

\begin{table*}
\begin{center}
\caption{Critical points for the models}
\begin{tabular}{lllll}
\hline
\multicolumn{1}{}
.System & $T_c$                  & ${\rho}_c$                     & $P_c$                                               & $C_f$ \\
\hline
V1     &  7.575 $\pm$ MeV        & 0.94$\pm 0.01 \times {\rho}_0$ & $(3.2 \pm 0.1){\times} 10^{-3}$ MeV$/$fm$^3$        & $0.49 \pm 0.02$\\
V2     &  6.875 $\pm$ MeV        & 0.74$\pm 0.01 \times {\rho}_0$ & $(7.3 \pm 0.1){\times} 10^{-4}$ MeV$/$fm$^3$        & $0.16 \pm 0.02$\\
V3     & 10.975 $\pm$ MeV        & 0.85$\pm 0.01 \times {\rho}_0$ & $(2.4 \pm 0.1){\times} 10^{-3}$ MeV$/$fm$^3$        & $0.28 \pm 0.02$\\
vdW    & 1.01   $\pm 0.01 \times T_c$ & 1.01$\pm 0.01 \times {\rho}_c$ & $(1.7 \pm 0.1){\times} 10^{-4}$  eV$/$A$\times V_c$ & $0.37 \pm 0.02$\\
\hline
\end{tabular}
\end{center}
\end{table*}

On first glance at Table 1 several noteworthy features stand out: (1) none of the values of $T_c$ determined for the models is the
same as the value of the parameter $T_c$ specified in the surface term in eq. (\ref{F_AZ_S}); (2) the critical densities for each 
model is close to unity; and (3) the critical temperature of V1, the model which includes the Coulomb force, is larger than the 
critical temperature of V2, the model with no Coulomb force.\ \ The error of the analysis of the van der Waals fluid was on the order 
of a few percent; $T^{vdW}_{c} \ne 1$.\ \ This illustrates the error inherent with this type of analysis.

It is not surprising that the critical temperature found by the analysis of thermodynamical quantities is different than the 
parameter $T_c$ used to parameterize the surface free energy of infinite nuclear matter.\ \ If one considers the critical point 
to be that temperature at which the surface free energy vanishes, then this can only be at $T_c$ (16 MeV in this case).\ \ However, 
the form of the surface term given by eq. (\ref{F_AZ_S}), approximately the form of the macroscopic surface free energy of a fluid 
near its critical point, is not an appropriate description of the microscopic surface of a droplet \cite{stauffer_kiang}.\ \ Moreover, 
eq. (\ref{F_AZ_S}) leads to a specific heat which approaches negative infinity as $T$ approaches $T_c$.\ \ In a more fundamental model 
the critical temperature would be an output of the model rather than an input.\ \ Of the phenomenological models studied here, only 
the van der Waals model is known to be self-consistent. 

The high densities found at the critical point for V1, V2, and V3 cannot be realized if one is constrained to placing spherical 
nuclei without overlaps inside of the total volume.\ \ Again this issue is mentioned but not dealt with since the aim of this 
paper is to explore only the thermodynamical predictions of the above models.

Were this sort of model interpreted physically, the high value of the critical density determined here would suggest that the 
critical point could never be reached by finite nuclear matter as a multiplicity of spherical clusters at normal nuclear density 
could not physically fit into the critical volume.\ \ If the constraint that the breakup volume must be large enough to avoid 
overlapping volumes of the final state (spherical) nuclei is added, then only first order phase transitions are possible.\ \ Of 
course there are several problems with a strictly physical interpretation of models such as the one presented here, not the least 
of which is the introduction of a volume for the system.\ \ Actual nuclei excited to high energies in nucleus-nucleus collisions 
do not exist in a box and thus have no volume in the sense suggested here.

Item (3), the rise in $T_c$ with the vanishing of the Coulomb force, is, on the surface, counterintuitive.\ \ Many other models 
of nuclear systems show just the opposite behavior \cite{jaqaman}, \cite{de}.\ \ However, those models are fundamentally different
than the ones examined in this work.\ \ Such models begin by describing the free energy or chemical potential or some equivalent 
quantity using formulae which assume a uniform distribution of material in much the same way that the van der Waals fluid assumes 
a uniform density.\ \ The model in this work samples a different part of the final state phase space \cite{bhattacharyya}.\ \ It will be 
seen that the calculation of the Coulomb term via eq. (\ref{E_AZ_C}) gives rise to the counterintuitive rise in the critical temperature 
when the Coulomb force is suppressed in the model.

To begin to understand the effect of the Coulomb energy on the critical point the same isotherm for models V1 and V2 was examined.\ \ 
Figures 4a and b show the isotherm of $T = 7.2$ MeV for V1 and V2.\ \ When the coexistence curve is shown for both systems, it is obvious 
that for V1 the isotherm is sub-critical while for V2 the isotherm is super-critical.\ \ Figures 4c and d begin to shed light on the cause 
of this counterintuitive occurrence.

\begin{figure} [ht]
\centerline{\psfig{file=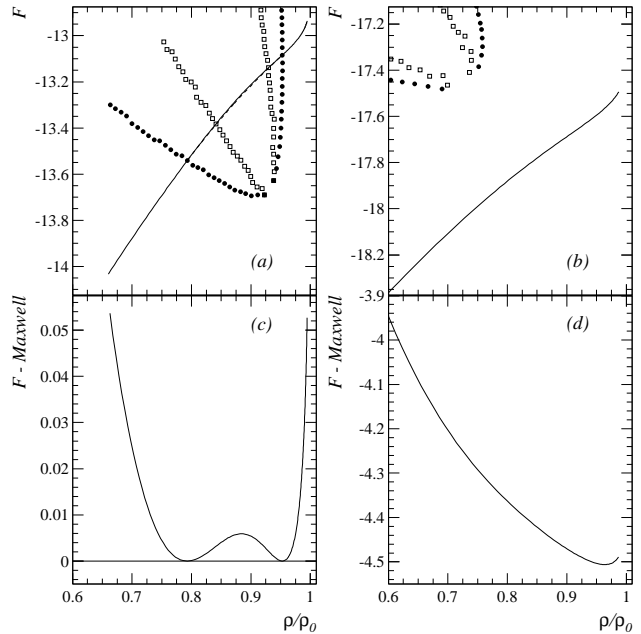,width=9.6cm,angle=0}}
\caption{Free energy isotherms as a function of reduced density for (a) V1 and (b) V2.\ \ For both systems the isotherm is for $T
= 7.2$ MeV.\ \ Open squares show the spinodal, full circles show the coexistence curve.\ \ The {\it Maxwell constructed} free energy
is shown as a dashed line through the coexistence region.\ \ The difference of the computed free energy and the {\it Maxwell
constructed free energy}  shows the presence of inflections points in (c) V1 at the coexistence boundary for this sub-critical
isotherm and none for (d) V2 for this super-critical isotherm.}
\label{fig:04}
\end{figure}

For the sake of illustration, the Maxwell constructed (a line of constant slope through the coexistence region, which then leads to a 
constant value of the pressure through the coexistence region that will give "equal areas" on a pressure-volume plot, see Figure 5a) 
path of the free energy through the coexistence region is shown as a dashed line, barely visible just below the isotherm in the coexistence 
region, in Figure 6a.\ \ The Maxwell constructed free energy is a straight line through the coexistence region.\ \ The constant slope of 
the Maxwell constructed free energy leads to a constant pressure for the system in the coexistence region.\ \ Because the path of Maxwell 
constructed free energy is very close to the path of the free energy of V1, it is difficult to see the difference in a plot such as shown 
in Figure 6a.\ \ A plot of the difference in the calculated free energy of V1 and the Maxwell constructed free energy shows what gives rise 
to the van der Waals loops in Figure 5a.\ \ See Figure 6c.\ \ There are two inflection points in the curve of the calculated free energy 
of V1, these are shown more clearly as the points in Figure 6c where the curve shows an ordinate value of zero.\ \ These plots for a 
canonical system's free energy are in the same spirit as plots for a microcanonical system's entropy \cite{gross1}.\ \ It is clear from 
Figures 4b and d that there are no similar inflection points in the free energy curve for V2.\ \ Therefore the isotherm of $T = 7.2$ MeV 
in V2 is super-critical while the very same isotherm is sub-critical for V1.

\begin{figure} [ht]
\centerline{\psfig{file=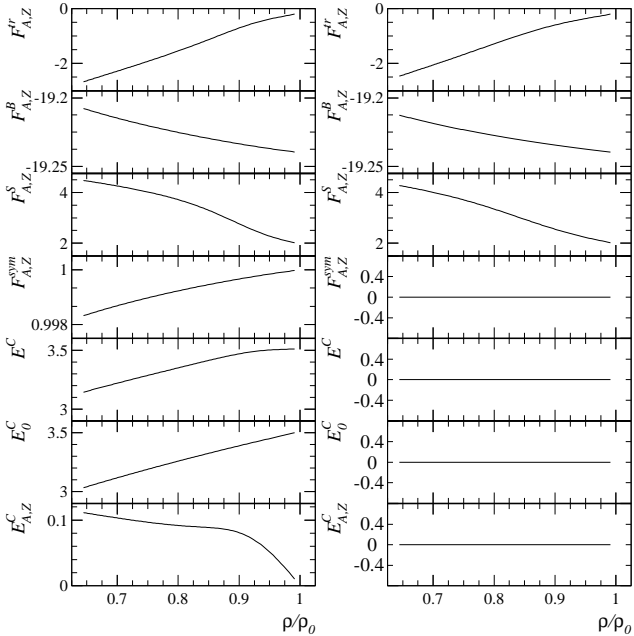,width=9.6cm,angle=0}}
\caption{Components of the free energy isotherms ($T = 7.2$ MeV) for V1 (left) and V2 (right).\ \ From top to bottom the components 
are for translation, bulk or Fermi, surface, symmetry, total Coulomb, uniform background Coulomb contribution and the clusterization
Coulomb contribution.\ \ Dotted lines show the coexistence boundary for this isotherm for V1}
\label{fig:05}
\end{figure}

It is possible to understand what gives rise to the inflection points introduced when going from V2 to V1 by looking at the
contributions to the total free energy of each system along the isotherm $T = 7.2$ MeV.\ \ See Figure 7.\ \ Plotted in this figure
for both systems are the components of the overall free energy for the isotherm in question: translational, bulk, surface, 
symmetry, total Coulomb, background Coulomb and clusterization Coulomb free energies.\ \ Figure 7 shows that the translational and 
bulk free energies of each system are nearly identical.\ \ An inspection of eqns (\ref{omega}) and (\ref{F_AZ_B}) shows that these
quantities are relatively insensitive to small changes in the fragment distribution.\ \ On the other hand, the surface free energy
shows an obvious difference in behavior between systems.\ \ In V1 the initial decrease in surface free energy as a function of
reduced density is slower than in V2.\ \ An inspection of eq. (\ref{F_AZ_S}) shows that the $A^{2/3}$-term introduces more
sensitivity to the fragment distribution than the previously discussed terms.\ \ The cause of this difference, and all the
differences between these two systems, is the presence of the Coulomb force in V1 and its absence in V2.

The behavior of the Coulomb contribution to the free energy of this isotherm is now discussed.\ \ The first, and simplest,
additional term in the free energy due to the Coulomb force is the asymmetry term.\ \ The change is in the asymmetry term is 
smooth as a function of increasing density and will not introduce the inflection points in the free energy curve that will change 
a super-critical isotherm to a sub-critical isotherm.

In this model the total Coulomb contribution to the free energy comes from two sources.\ \ One which represents the background
energy due to a uniform distribution of charges, eq. (\ref{coulomb}), and the other due to the energy from the clusterization of
fragments, eq. (\ref{E_AZ_C}) \cite{bondorf}.\ \ The background energy, $E_{0}^{c}$, goes as ${\rho}^{1/3}$ and therefore varies 
smoothly with volume.\ \ The clusterization free energy, $E^{C}_{A,Z}$, shown in Figure 7 has a different behavior.\ \ It decreases 
as the reduced density increases, the rate of decrease is at first nearly constant, then slows and then increases rapidly over some 
small interval in reduced density.\ \ The combination of these two Coulomb terms introduces sufficient changes in the overall free 
energy of the system from V2 to V1 that inflection points arise and thus the critical temperature increase when the Coulomb force is 
added to the system.\ \ A smooth or constant version of the Coulomb free energy added to version V2 should not cause this sort of
behavior.\ \ It is the variation in $E^{C}_{A,Z}$ that introduced the inflection points and increases the critical temperature.\ \
In the end, it is the behavior of the free energy curve that served to determine the location of the critical point and that
behavior is, at times, counter intuitive.

Also listed in Table I is the compressibility factor:
\begin{equation}
	C_f = \frac{P_c V_c}{T_c} .
\label{compfact}
\end{equation}
The text book value for $C_f$ for the van der Waals gas is recovered to within error bars.\ \ According to the law of corresponding 
states, the value of $C_f$ should be universal.\ \ For fluid systems this is the case and $\left< C_f \right> \sim 0.292$ 
\cite{gugg_text}.\ \ For V1 and V2 there no such universal behavior observed to within error bars, while V2 shows some degree of 
universality.

\subsection{Isochores}

The volume of the system is held constant and the behavior of various quantities with respect to the system's temperature is
explored.

\begin{figure} [ht]
\centerline{\psfig{file=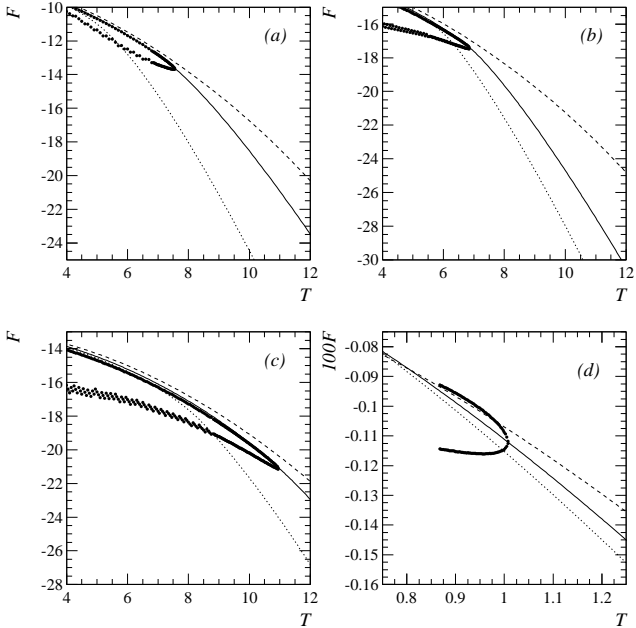,width=9.6cm,angle=0}}
\caption{Free energy isochores for (a) V1, (b) V2, (c) V3 and (d) the van der Waals fluid.\ \ Full circles show the boundary of the
coexistence region.\ \ Dotted curves are sub-critical.\ \ Solid curves are critical.\ \ Dashed curves are super-critical.}
\label{fig:06}
\end{figure}

Beginning again with the primary quantity calculated, Figure 8 shows the free energy for each system as a function of the system's
temperature.\ \ For models V1, V2 and V3 the isotherms are for $\rho = {\rho}_0 / 3$, $\rho = {\rho}_c$, and $\rho = 0.9995
{\rho}_0$.\ \ For the van der Waals fluid, the isochores shown are for $\rho = 1.25 {\rho}_c$, $\rho = {\rho}_c$ and $\rho = 0.75
{\rho}_c$.\ \ See table I for critical density values.\ \ Also shown are the values of the free energy of the systems along their 
respective coexistence curves.

In reference \cite{mdgm} the authors show results of a similar model similar to V2, for an isochore of approximately ${\rho} =
{\rho}_0 / 3$.\ \ In that work there was a {\it kink} in the free energy curve which was interpreted as evidence for a first
order phase transition.\ \ For the small system used in this work the {\it kink} is smoothed out negating the efficacy in using 
the kink as evidence towards determining the order of a phase transition if one is present.\ \ In larger systems, systems in ref. 
\cite{mdgm} were more than 15 times larger than the system used in this work, the kink is more evident and this procedure may be 
possible.

Knowledge of the location of the coexistence curve allows for the identification of sub-critical, critical and super-critical
isochores.\ \ For the projections shown in Figure 8 all the versions of the nuclear model a sub-critical isochore crosses the 
coexistence region.\ \ The critical isochore travels along the high value edge of the coexistence curve passing through the critical 
point.\ \ Super-critical isochores do not come into contact with the coexistence curve nor do they traverse the coexistence region.\ \ 
The behavior of the van der Waals fluid is somewhat different from the behavior of the nuclear models.\ \ Figure 8d shows that for a 
van der Waals fluid in this projection of the phase diagram all are in the coexistence region for low temperatures and cross the 
coexistence curve at higher temperatures.\ \ This apparent behavior results from projection of the three dimensional phase diagram 
onto a two dimensional plot.\ \ In a three dimensional figure, the super-critical isochore is seen to travel outside the coexistence 
region, the critical isochore is observed to intersect with the critical point and the sub-critical isochore is seen to traverse the 
coexistence region.\ \ See Figure 9.

\begin{figure} [ht]
\centerline{\psfig{file=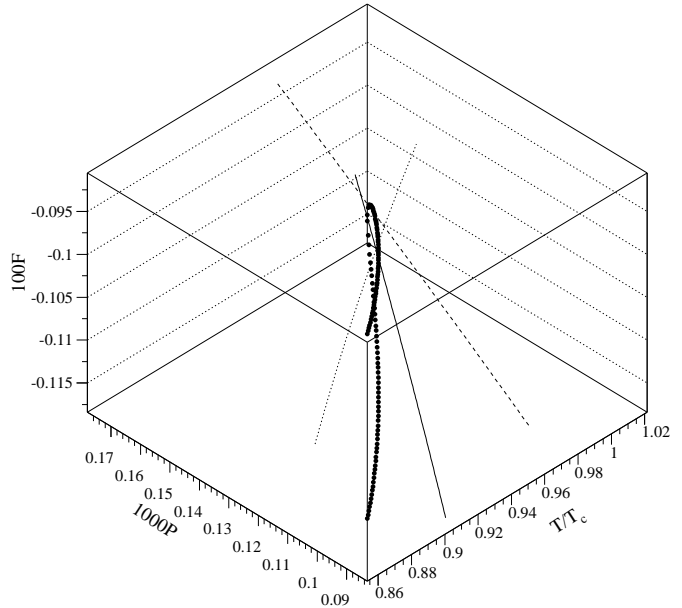,width=9.6cm,angle=0}}
\caption{A three dimensional plot of the free energy isochores for the van der Waals fluid.\ \ Full circles show the boundary of
the coexistence region.\ \ Dotted curves are sub-critical.\ \ Solid curves are critical.\ \ Dashed curves are super-critical.}
\label{fig:07}
\end{figure}

The pressure isochore was obtained by following the procedure outlined in eq. (\ref{pressure}) and using $\left<{\rho}\right>$, 
the average density, and $\left<T\right>$, the average temperature over the ${\Delta}T$ interval.\ \ See Figure 10.\ \ In the 
limit of vanishing ${\Delta}T$ this procedure is valid.\ \ Figure 10 shows the same three isochores discussed above as well as the 
location of the coexistence curve, seen edge on in this projection of the phase diagram.\ \ All the figures for the nuclear model 
show similar behaviors for the critical and super-critical isochore.\ \ All super-critical isochores follow a trajectory above the 
coexistence curve.\ \ The critical isochore for all versions of the model approaches the coexistence curve, follows it, and leaves 
at the termination or critical point.\ \ The sub-critical isochores for V2 and V3 pass through the coexistence curve as the 
temperature is increased and the system makes a first order transition from a liquid to gaseous state.\ \ For version V1 the behavior 
is more complicated.\ \ The sub-critical isochore begins on the gaseous side of the coexistence curve, crosses the coexistence curve 
into the liquid region before crossing back over the coexistence curve into the gaseous region at higher temperature.\ \ The $P$-$T$ 
projection of the van der Waals fluid looks as expected.\ \ See Figure 8d.

\begin{figure} [ht]
\centerline{\psfig{file=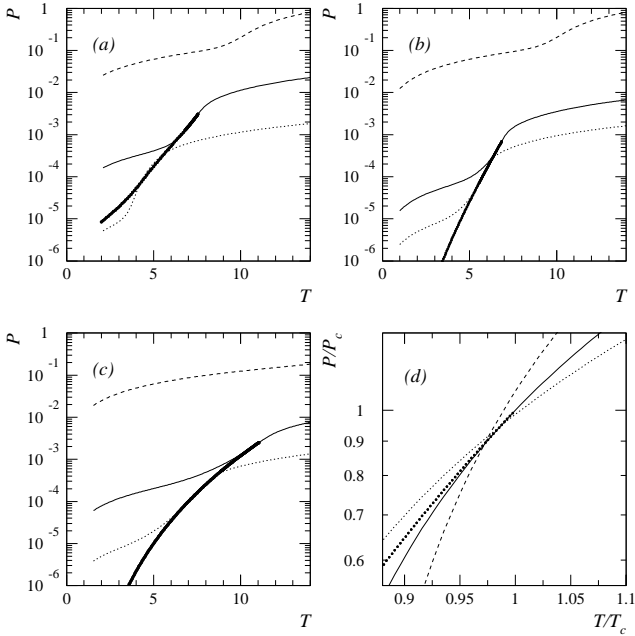,width=9.6cm,angle=0}}
\caption{Isochores of pressure versus temperature for (a) V1, (b) V2, (c) V3 and (d) the van der Waals fluid.\ \ Full circles show
the boundary of the coexistence region.\ \ Dotted curves are sub-critical.\ \ Solid curves are critical.\ \ Dashed curves are
super-critical.}
\label{fig:08}
\end{figure}

Following the analysis procedure outlined in the previous section, the entropy was determined via eq. (\ref{entropy2}).\ \ See
Figure 11.\ \ The same three isochores are plotted showing the entropy as a function of temperature.\ \ All three isochores for 
the nuclear models show a smooth rise as a function of temperature with some region of increased slope.\ \ This behavior is 
consistent with a continuous phase transition, if a phase transition were present.\ \ Along an isochore, a first order transition 
would be indicated by a sudden change the behavior of the entropy as a function of temperature at one edge of the coexistence 
region.\ \ However, due to the small size of the system such sharp behavior is smoothed out into smooth curves making it 
impossible to draw a conclusion about the order of a phase transition from plots such as those shown in Figure 11.\ \ As before, 
knowing the location of the coexistence curves makes identification sub-critical, critical and super-critical isochores in the 
nuclear model trivial.\ \ The sub-critical isochore traverses the coexistence region, the critical isochore passes through the 
critical point and the super-critical isochore avoids the coexistence region.\ \ Also as before the van der Waals system shows a 
different behavior, and an added dimension to the plot in Figure 11d must be made to understand the behavior of the various 
isochores.

\begin{figure} [ht]
\centerline{\psfig{file=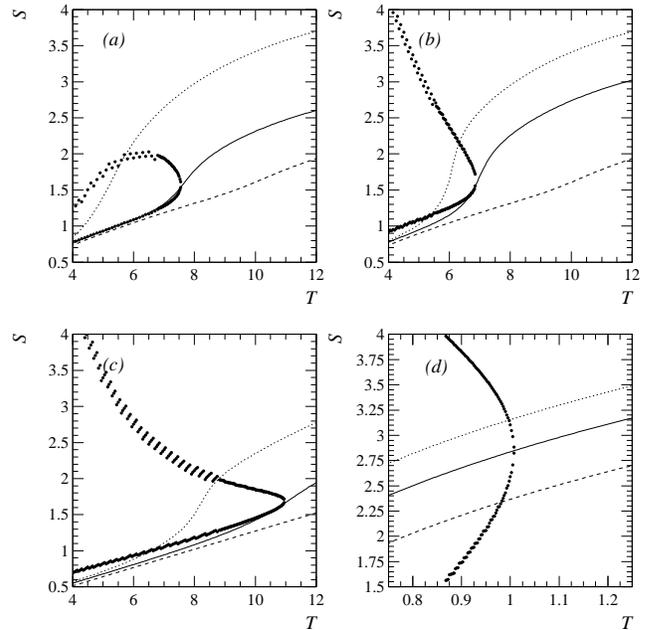,width=9.6cm,angle=0}}
\caption{Isochores of the entropy versus temperature for (a) V1, (b) V2, (c) V3 and (d) the van der Waals fluid.\ \ Full circles
show the boundary of the coexistence region.\ \ Dotted curves are sub-critical.\ \ Solid curves are critical.\ \ Dashed curves are
super-critical.}
\label{fig:09}
\end{figure}

The specific heat at a constant volume, $C_V$, for each system is shown in Figure 12, again for the same three isochores.\ \ All 
curves show a peak in that could be due to a smoothed out discontinuity (first order phase transition), the remnants of a power law
divergence at the critical point (continuous phase transition) or a specific heat anomaly (super-critical behavior).\ \ As with the
free energy and entropy it is impossible to come to a definite conclusion regarding the presence and nature of a phase transition
from this plot.\ \ Of particular importance is the specific heat for the van der Waals fluid which is seen to be nearly constant
and equal to $3 / 2$ as it should \cite{goldenfeld}.\ \ There is a small slope for the specific heat over the temperature range in
question: for $T = T_c / 2$ $C_V = 1.507$ and for $T = 2 T_c$ $C_V = 1.502$.\ \ This illustrates the technique employed here,
following equations (\ref{entropy2}) and (\ref{cv}), yields results that are no better than $0.25$\% for the quantities determined
in this paper.

\begin{figure} [ht]
\centerline{\psfig{file=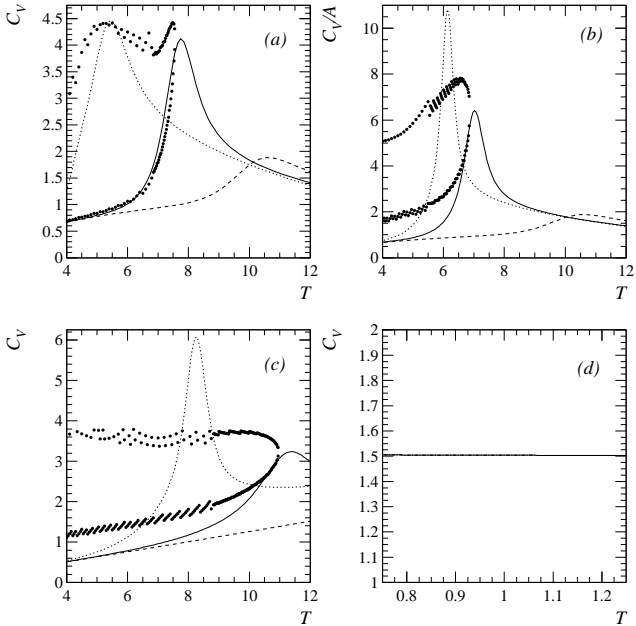,width=9.6cm,angle=0}}
\caption{The specific heat at constant volume versus temperature for (a) V1, (b) V2, (c) V3 and (d) the van der Waals fluid.\ \
Full circles show the boundary of the coexistence region.\ \ Dotted curves are sub-critical.\ \ Solid curves are critical.\ \
Dashed curves are
super-critical.}
\label{fig:10}
\end{figure}

Finally, Figure 13 shows the constant volume caloric curves of $T$ as a function of $E$ for each system at the same isochores
discussed above.\ \ Also shown are the values of the temperature and energy of the system, $E = F + TS$, along the boundary of the
coexistence region.\ \ Just as with the behavior of the entropy isochores, the behavior of the caloric curves for each isochore
shows behavior that is consistent with either a first order or continuous phase transition in a small system.\ \ Each isochore
shows similar behavior, a steep rise followed by a region of shallower incline followed by a portion which approaches $E/A =
\frac{3}{2} T$.\ \ The lack of a flat region, or back bend, in the caloric curve is not due to the small size of the system 
but rather due to the system being held at a constant volume.\ \ Only for {\it isobars} will flat regions or back bends be 
observed in the canonical nuclear model.\ \ See following section.\ \ Again the van der Waals system shows very different 
behavior, a steady rise in the temperature as a function of energy.\ \ And again a three dimensional plot is needed to clearly 
understand the nature of each isochore.

\begin{figure} [ht]
\centerline{\psfig{file=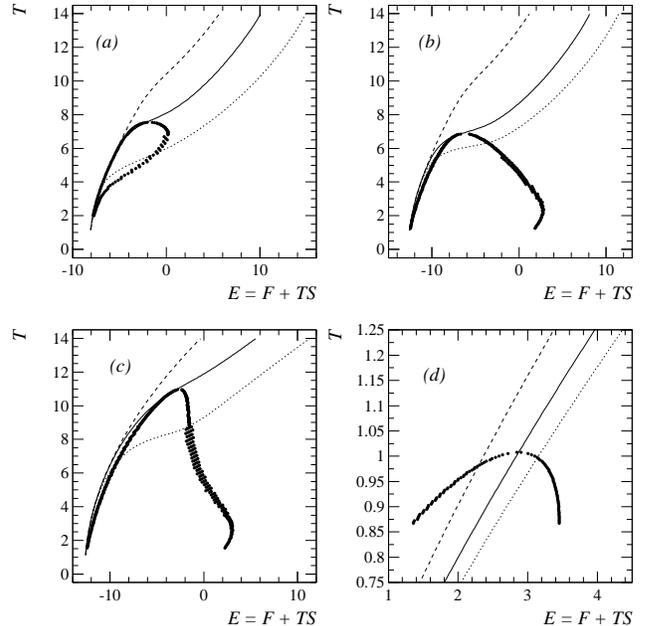,width=9.6cm,angle=0}}
\caption{Isochoric caloric curves for (a) V1, (b) V2, (c) V3 and (d) the van der Waals fluid.\ \ Full circles show the boundary of
the coexistence region.\ \ Dotted curves are sub-critical.\ \ Solid curves are critical.\ \ Dashed curves are super-critical.}
\label{fig:11}
\end{figure}

\subsection{Isobars}

The pressure of the system is held constant and the behavior of various quantities with respect to the system's temperature is
explored.\ \ Here some care should be taken with the interpretation of the results.\ \ The analysis outlined above is still
followed.\ \ However, since the pressure is an extensive quantity plots such as free energy versus temperature are now plots of
$\left<F\right>$ versus $T$.\ \ Where $\left<F\right>$ is the mid point in the ${\Delta}F$ range over which a difference such as 
eq. (\ref{pressure}) is taken.\ \ In the limit of vanishing interval size, this approximation is accurate.\ \ Also because the 
pressure is an extensive variable, it was necessary to allow some small variation in $P$ in order to make a plot such as $\left< 
F \right>$ against $T$.\ \ The variation in $P$ was usually less than one percent; {\it i.e.} $\Delta P / P \leq 0.01$.\ \ Small 
changes in the amount of variation of $P$ had little effect on the analysis presented here.\ \ Large changes in the variation in 
$P$ wash out the behavior observed below.

The isobaric free energy as a function of temperature is shown for all systems in Figure 14 as well as the values of the isobaric
free energy along the boundaries of the coexistence region.\ \ In all systems there is a back bend in the free energy curve for
sub-critical isobars.\ \ The sub-critical isobar also traverses the coexistence region.\ \ It would be possible to perform the
Maxwell construction procedure and deduce the critical point from these plots.\ \ The van der Waals fluid system shows that the
critical point determined in the construction of the $P$-$V$ coexistence curves agrees with considerations of isobaric $F$.\ \ See 
Figure 14d.\ \ The critical isobar shows a vertical slope tangent to the coexistence curve and no back bend.\ \ The super-critical 
isobar does not traverse the coexistence region and shows no back bend.

\begin{figure} [ht]
\centerline{\psfig{file=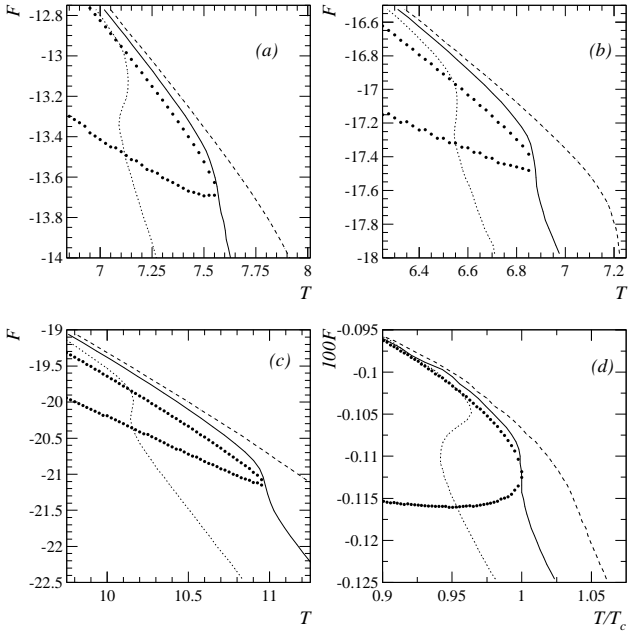,width=9.6cm,angle=0}}
\caption{Free energy isobars as a function of the temperature for (a) V1, (b) V2, (c) V3 and (d) the van der Waals fluid.\ \ Full
circles show the boundary of the coexistence region.\ \ Dotted curves are sub-critical.\ \ Solid curves are critical.\ \ Dashed
curves are super-critical.}
\label{fig:12}
\end{figure}

Figure 15 shows the isobaric temperature as a function of reduced density.\ \ Temperature is plotted as a function of reduced 
density in the spirit of the Guggenheim plot \cite{guggenheim} which shows the universal behavior of several fluids near their 
critical point.\ \ On the scales shown here no universal behavior is observed.\ \ It may be that very near the critical point, the 
coexistence curves for each system are identical.\ \ The $T$-$\rho$ isobars for all systems show the expected behaviors: a sub-critical 
back bending curve that traverses the coexistence region giving way to the critical isobar, a critical curve with a flat section which 
intercepts the coexistence region at the critical point and finally a super-critical isobar which avoids the coexistence region 
altogether.

\begin{figure} [ht]
\centerline{\psfig{file=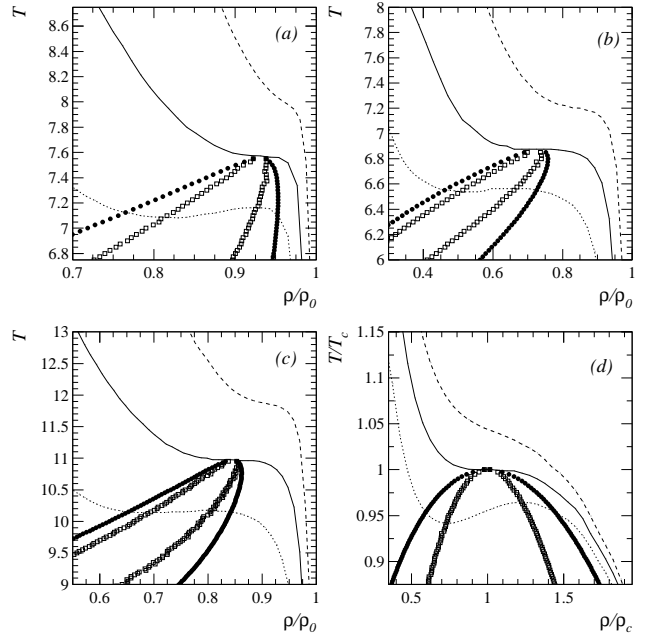,width=9.6cm,angle=0}}
\caption{Isobars of temperature versus reduced density for (a) V1, (b) V2, (c) V3 and (d) the van der Waals fluid.\ \ Open 
squares show the spinodal.\ \ Full circles show the boundary of the coexistence region.\ \ Dotted curves are sub-critical.\ \ 
Solid curves are critical.\ \ Dashed curves are super-critical.}
\label{fig:13}
\end{figure}

The behavior of the isobaric entropy as a function of temperature is also just as expected.\ \ See Figure 16.

\begin{figure} [ht]
\centerline{\psfig{file=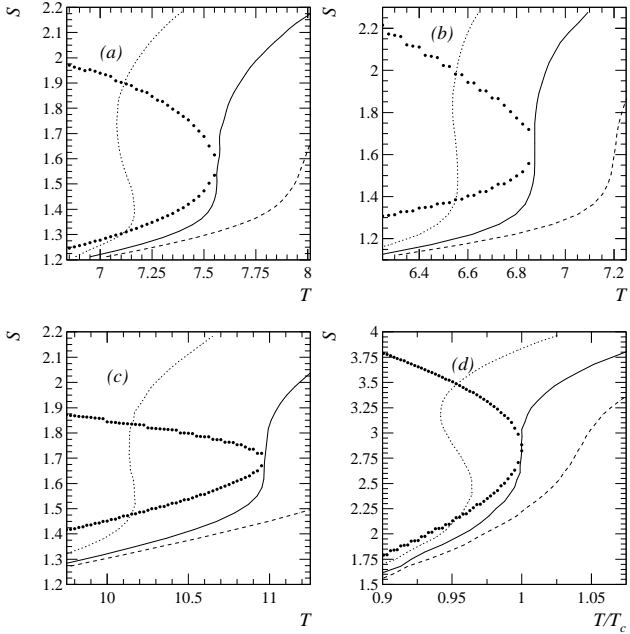,width=9.6cm,angle=0}}
\caption{Isobars of the entropy versus temperature for (a) V1, (b) V2, (c) V3 and (d) the van der Waals fluid.\ \ Full circles show
the boundary of the coexistence region.\ \ Dotted curves are sub-critical.\ \ Solid curves are critical.\ \ Dashed curves are
super-critical.}
\label{fig:14}
\end{figure}

Figure 17 shows the isobaric caloric curves for each system.\ \ As the energy is constructed from the free energy and entropy of
the system the back bending observed in the sub-critical isobars is expected.\ \ Also shown is the value of temperature and energy
along the boundary of the coexistence region.\ \ Note the difference between the isochoric caloric curves shown in Figure 13 and
the isobaric caloric curves shown in Figure 17.\ \ When the pressure is held constant all of the canonical systems discussed here,
including the van der Waals fluid, show sub-critical caloric curves with a back bend.\ \ No back bending is present when the volume
is held constant.

\begin{figure} [ht]
\centerline{\psfig{file=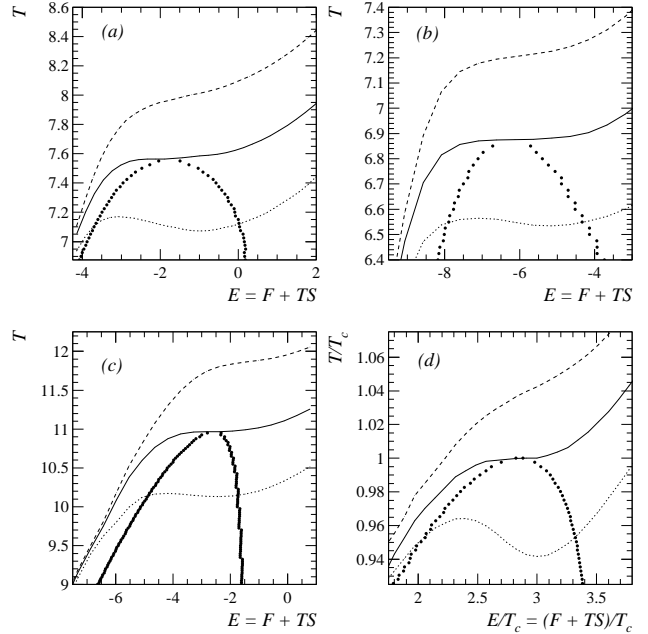,width=9.6cm,angle=0}}
\caption{Isobaric caloric curves for (a) V1, (b) V2, (c) V3 and (d) the van der Waals fluid.\ \ Full circles show the boundary of
the coexistence region.\ \ Dotted curves are sub-critical.\ \ Solid curves are critical.\ \ Dashed curves are super-critical.}
\label{fig:15}
\end{figure}

Finally Figure 18 shows the constant pressure specific heat, $C_P$, as calculated from:
	\begin{equation}
		C_P = \left( \frac{{\Delta} E}{{\Delta} T} \right)_{P} 
	\label{cp}
	\end{equation}
taking as input the isobars shown in Figure 17.\ \ The results for $C_P$ of these canonical calculations are similar to the
behavior reported in micro canonical models for various systems \cite{gross}, \cite{gross1}, \cite{wales1}, \cite{wales2}.\ \ 
The sub-critical isobars show the remnants of poles with $C_P < 0$ values in between.\ \ The critical isobar shows the remnants 
of a divergence, and the super-critical isobar shows some peaking behavior.\ \ The lack of poles and divergences is not 
due to the finite size of the systems V1, V2 and V3, but rather due to the computational nature of these calculations.\ \ The 
calculations for the van der Waals fluid are, in effect, for a truly thermodynamic system and the van der Waals critical isobar 
of $C_P$ still shows no true divergence.\ \ The critical isobar of $C_P$ for V2 shows a negative value which is due to the 
computational nature of the calculation and the manner in which $C_P$ was calculated from $E$ and $T$ values.

\begin{figure} [ht]
\centerline{\psfig{file=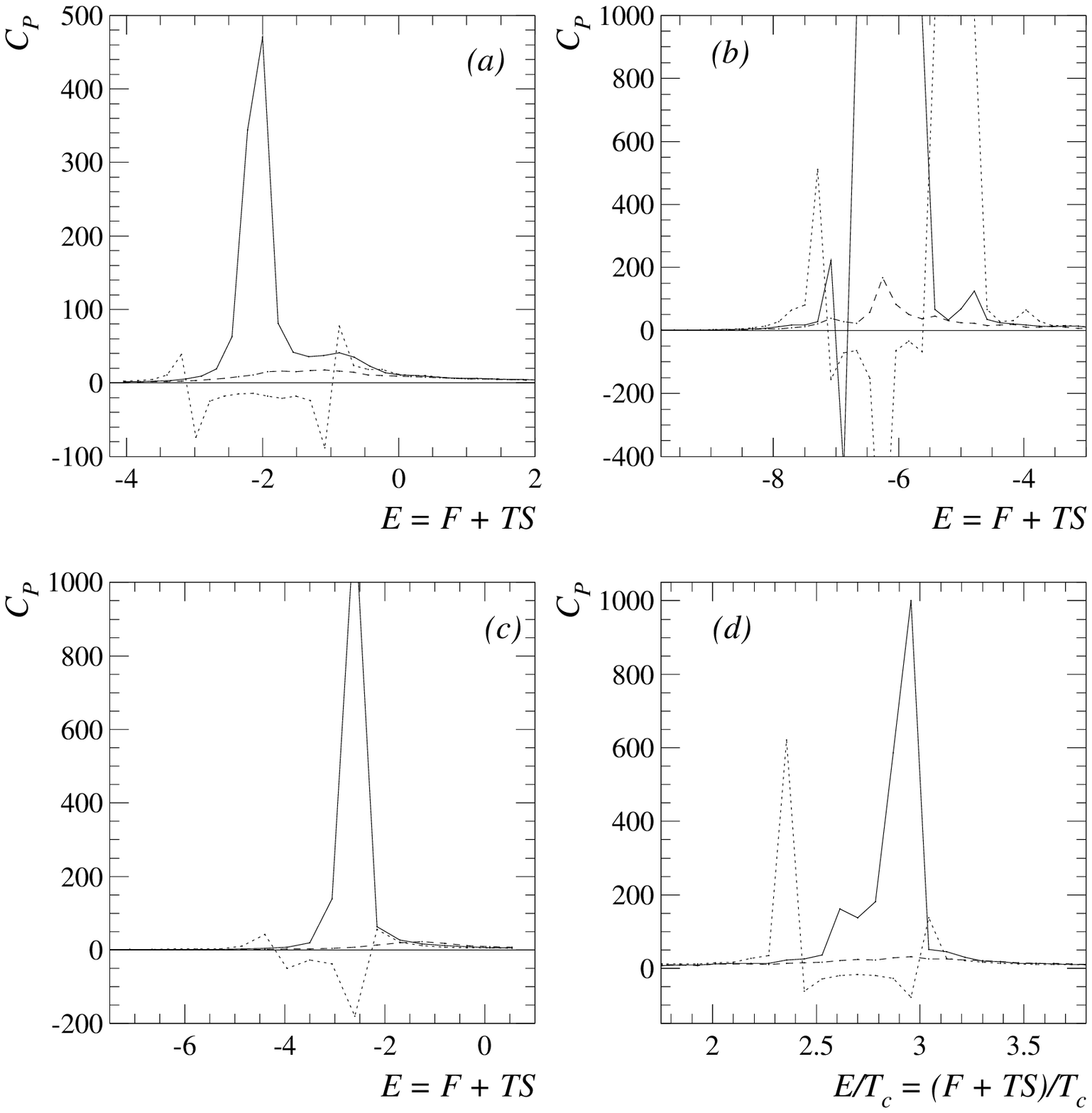,width=9.6cm,angle=0}}
\caption{Specific heat at a constant pressure versus energy for (a) V1, (b) V2, (c) V3 and (d) the van der Waals fluid.\ \
Dotted curves are sub-critical.\ \ Solid curves are critical.\ \ Dashed curves are super-critical.}
\label{fig:16}
\end{figure}

\subsection{Iso-nothing: Variable $(P, V, T)$}

The nuclear model presented here assumes a system enclosed in some volume.\ \ An actual excited nucleus is not enclosed in a volume.\ \ 
This has the effect of forcing a path through the thermodynamic phase space which is considerably different from any of the paths 
investigated thus far.\ \ To bridge the gap between reality and tractable calculations, an energy dependent free volume is assumed in 
models such as the ones presented in this work \cite{bondorf}.\ \ At low energies the free volume of the system is assumed to be nearly 
constant and vanishingly small.\ \ At a given energy the system is allowed to expand and the free volume increases from near zero.\ \ 
This has the effect of tracing a path through the thermodynamic phase space of the system off any of the trivial paths along one of the 
axes investigated above.\ \ It is possible to examine the effects of such a parameterization of the free volume as a function of energy 
with the calculations made here.\ \ Calculations in this section were performed only for V1.

Following the same ideas presented in the discussion of isobars, the values of $F$, $S$ and $T$ are determined along a path in 
$V_f$ as a function of $E$.\ \ As $V_f$ changes values of $F$, $S$ and $E$ are picked from the appropriate isochore.\ \ For example, 
instead of traveling along a path parallel to one of the axes in Figure 2a, {\it e.g.} an isochore, isotherm or isobar, an energy 
dependent free volume was chosen so that the system evolved through thermodynamic phase space on a non-trivial trajectory.\ \ See 
Figure 19.\ \ In Figure 19a small points show a small sample of the set of calculations for $F(T,V)$.\ \ Larger points show the 
values of $F(V,T)$ selected for a $V_f (E)$ trajectory described above.\ \ Figures 19b and c show the same for $S(T,V)$ and $E(T,V)$.

\begin{figure} [ht]
\centerline{\psfig{file=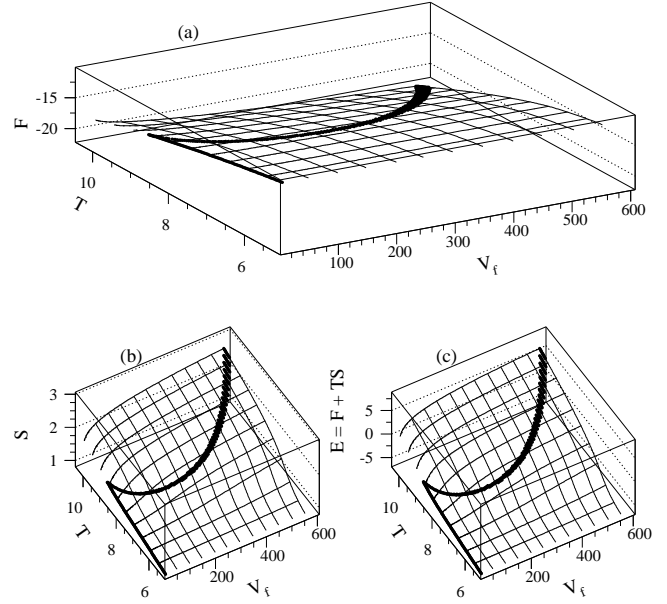,width=9.6cm,angle=0}}
\caption{Solid circles show the path of an expanding nuclear system, V1, through thermodynamic phase space.\ \ Neither volume nor 
pressure is held fixed.\ \ Points show the calculations of the free energy surface as shown in Figure 3.} 
\label{fig:17}
\end{figure}

For the purposes of the present analysis, values of $S(T,V)$ were used from different isochores: as $V_f$ changed values of $S(T,V)$ 
were selected from the appropriate isochore.\ \ If the change from isochore to isochore is small, ${\Delta}V_f \sim 0$, then the 
procedure is a good approximation.\ \ The intervals in $V_f$ for the calculation of $F(T,V)$ for this analysis were on the order 
of two percent of the free volume of the system; {\it e.g.} when $V_F = 500$fm$^3$, ${\Delta}V_f \sim 10$fm$^3$ and when 
$V_f = 5$fm$^3$, ${\Delta}V_f \sim 0.1$fm$^3$.

Figure 20a shows three different paths through thermodynamic phase space: one which travels very near to the critical point and 
into the coexistence region (solid curve), and two others which avoid the coexistence region all together (dotted and dashed 
curves).\ \ As mentioned previously for the full version of this model \cite{bondorf} the free volume is nearly zero for 
the lower end of the energy range.\ \ At some energy, $2$ MeV/A for the solid curve in Figure 20, the system is allowed to expand 
and the free volume increases as a function of energy.\ \ The functional form of $V_f (E)$ is not identical to other models 
\cite{bondorf} but close enough to show the same behavior observed in the full version of the model \cite{bondorf}.\ \ Also shown 
in Figure 20a are the values of the free volume and energy along the boundary of the coexistence region.\ \ Note that for the solid 
curve the $V_F (E)$ trajectory enters the coexistence region near the critical point and leaves the coexistence region at a higher 
energy and free volume.\ \ The other two trajectories shown in Figure 20 have the same general behavior: increasing free volume 
with increasing energy, but the precise paths differ.

\begin{figure} [ht]
\centerline{\psfig{file=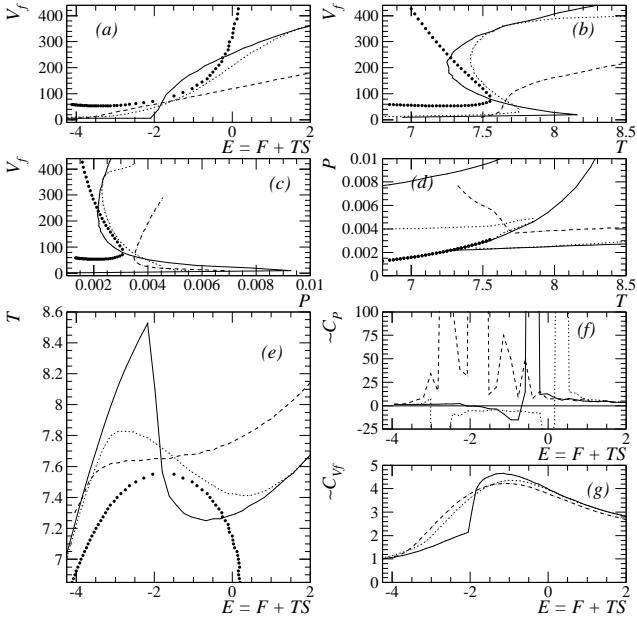,width=9.6cm,angle=0}}
\caption{The path of an expanding nuclear system, V1, through thermodynamic phase space: (a) the free volume as a function of energy, 
(b) free volume as function of temperature, (c) free volume as a function of pressure, (d) pressure as function of temperature, (e) 
the resulting caloric curve, (f) a specific heat based on the derivative of the caloric curve and (g) a specific heat along the phase
space trajectory.\ \ Full circles show the boundary for the coexistence region.\ \ Different line types show different $(P, V, T)$ 
trajectories.\ \ See text for details.}
\label{fig:18}
\end{figure}

In a plot of free volume against temperature back bending is observed for two of the trajectories presented here.\ \ See Figure 
20b.\ \ The solid curve trajectory of $V_f (T)$ begins with a small free volume that is constant until a temperature of just over 
$8$ MeV, then the system expands and cools.\ \ This is shown by the back bend.\ \ The solid curve $V_f (T)$ trajectory then enters 
the coexistence region near the critical point.\ \ At a temperature between $7$ and $7.5$ MeV the slope of the free volume nearly 
diverges and then changes in sign.\ \ After this point, further expansion in the free volume is accompanied by an increase in 
the temperature of the system.\ \ The $V_f (T)$ curve then leaves the coexistence region at $T < T_c$ and $V > V_c$.\ \ The other 
trajectories show similar, but less extreme behavior.

Other projections of the phase diagram for this model with the solid curve $V_f (E)$ trajectory show back bends as well.\ \ See
Figures 20c and d.\ \ The $V_f$-$P$ projections shows the system's pressure increases nearly an order of magnitude over the
constant free volume section.\ \ When the system is allowed to expand, the pressure drops.\ \ The trajectory passes near the
critical point as it enters the coexistence region and in the course of back bending exits the coexistence region.

The solid curve $P$-$T$ trajectory is equally interesting.\ \ As the system increases in temperature with a fixed free volume, the 
pressure increases.\ \ When the system reaches a temperature between $8$ and $8.5$ MeV the expansion sets in and the pressure and 
temperature both drop so that the trajectory moves towards the critical point.\ \ At a temperature between $7$ and $7.5$ MeV, the 
system reverses the trend and both pressure and temperature increase.\ \ See Figure 20d.\ \ Again the other two trajectories show 
similar behavior, but to a lesser extent as the $V_f (E)$ trajectory becomes smoother.

As back bending has already been observed for other $V_f (E)$ trajectories, it is no surprise that the caloric curve for this
changing volume system also shows a back bend.\ \ See Figure 20e.\ \ The solid caloric curve shown here is reminiscent of other back
bending caloric curves already published in reference \cite{bondorf}, \cite{bondorf1} - \cite{agostino} where variable free 
volume constrained canonical and constrained grand canonical calculations are made, but not those published in reference 
\cite{gross}, \cite{gross1} where a constant volume micro canonical calculation is made.\ \ Also shown in Figure 20e are the 
values of the temperature and energy along the coexistence curve.\ \ Again the trajectory of the solid curve variable free volume 
enters the coexistence region near the critical point and exits the coexistence region at $T < T_C$ and $E > E_c$.\ \ The dotted 
caloric curve also shows back bending, albeit to a more limited extent and the dashed caloric curves shows no major back bend.

Next the specific heat of the system was calculated via eq. (\ref{cp}).\ \ See Figure 20f.\ \ The application of eq. (\ref{cp}) to
this trajectory through thermodynamic phase space is problematic.\ \ From Figure 20d it is clear that pressure is not a constant 
and thus eq. (\ref{cp}) should not be used.\ \ However, it has become commonplace to follow this sort of procedure \cite{bondorf1},
\cite{agostino} even though it is in contradiction with definition of $C_P$ or $C_V$.\ \ When eq. (\ref{cp}) is applied 
to the solid and dotted curves in Figure 20e, the resulting specific heat shows negative values and the remnants of a divergence.

Finally the specific heat of the system was determined in the same manner that the entropy of the system was determined.\ \ The
specific heat at a constant volume was calculated along an isochore, then the values of $C_V$ were selected from the
paths through thermodynamic phase space of the $V_f (E)$ trajectory.\ \ See Figure 20g.\ \ For the solid like path, the value of 
$C_V$ shows a steady rise until an energy of $2$ MeV/A and then a sharp rise as the system expands.\ \ As the system continues to 
expand and the energy increases the value of $C_V$ reaches a maximum and then shows a gradual decline.\ \ No $C_V < 0$ is observed 
in this plot.\ \ The other two paths show smoother behavior.

The question that now arises is what, if any, insight into the nature of the phase transition can be obtained from curves such as 
those in Figure 20e and f.\ \ Were there no other knowledge of the system, the back bends observed in the solid and dotted caloric 
curves would suggest that the system had gone through a first order phase transition.\ \ Negative values and a peak in the specific 
heat would seem to confirm this.\ \ While for the dashed curve, the lack of back bending in the caloric curve and the lack of a 
negative specific hear would argue either for a continuous phase transition of no phase transition.\ \ However, from the 
analysis of the previous sections, the location of the critical point and the shape and location of the coexistence curve are 
known.\ \ The addition of this knowledge makes it clear that the naive analysis of Figures 20e and f can provide misleading 
results.\ \ While the solid caloric curve shows a back bend, the trajectory of the systems goes through the critical point, into 
the coexistence region and exits along a sub-critical path.\ \ Can one conclude that the system has undergone a continuous phase 
transition, or has it undergone a first order phase transition because the trajectory traverses the coexistence region?\ \ The 
answer to the first question is yes, since the system does reach the critical temperature and density simultaneously.\ \ The answer 
to the second question also appears to be ``yes'' since the specific heat found from eq. (\ref{cp}) is less than zero (for the
solid curve).\ \  Furthermore, the solid curve intersects the coexistence curve at a $T$ less than $T_c$.\ \ Note, however, that 
without knowledge of the location of the coexistence curve, the above questions cannot be unambiguously answered.\ \ A naive inspection 
of the dotted caloric curve may lead one to conclude that the back bending is indicative of a first order phase transition. Such is not 
the case, Figure 20e shows that the dashed line never traverses the coexistence region.\ \ It is clear that drawing conclusions based 
on caloric curves is difficult unless one has knowledge of the complete thermodynamics of the system.

\subsection{Critical exponents from thermodynamic quantities}

Any model that attempts to describe a system capable of undergoing a continuous phase transition should exhibit quantities with
singular behavior that, near the critical point, are described by power laws with a consistent set of critical exponents.\ \ These
critical exponents should obey well known scaling laws and may, or may not, fall into one of the established universality classes.\
\ To that end, four critical exponent values are determined and three scaling laws are checked for these models using the critical
point, $( T_c , P_c , {\rho}_c )$, determined previously and other thermodynamic quantities.\ \ Note that in the determination of
critical exponents presented here, thermodynamic variables are used explicitly, {\it e.g.} in the extraction of $\gamma$ it is the
isothermal compressibility that is used, and not moments of the fragment distribution.

\subsubsection{Power law results}

The exponent ${\alpha}$ is determined by the behavior of the specific heat along the critical isochore, see Figure 12.\ \ The $C_V
(T)$ curve was fit with the functional form:
	\begin{equation}
		C_V (T) = H_{\pm} \left( \frac{ \left| \frac{T-T_c}{T_c} \right|^{-{\alpha}_{\pm}} - 1 }{\alpha_{\pm}} \right) 
                        + G_{\pm}
	\label{alpha}
	\end{equation}
on both sides of the critical point $T {\frac{>}{<}}T_c$ \cite{heller}.\ \ The fit parameters $H_{\pm} $, ${\alpha_{\pm}}$ and
$G_{\pm}$ were allowed to vary to minimize the ${\chi}^2$ of the fit.\ \ Figure 21 shows the results for each system and Table II
lists the extracted exponents.

\begin{table*}
\begin{center}
\caption{Critical exponents for the models}
\begin{tabular}{lllllll}
\hline 
System & ${\alpha}_{\pm}$ & ${\beta}$       & ${\gamma}$        & ${\delta}_{+}$  & ${\delta}_{+}$  & $\tau$          \\
\hline
V1     & $0.6 {\pm} 0.2$  & $0.5 {\pm} 0.2$ & $1.0  {\pm} 0.2$  & $1.2 {\pm} 0.3$ & $1.4 {\pm} 0.3$ & $2.49 \pm 0.03$ \\
V2     & $0.9 {\pm} 0.2$  & $0.6 {\pm} 0.1$ & $1.0  {\pm} 0.4$  & $2.8 {\pm} 0.1$ & $2.7 {\pm} 0.2$ & $2.10 \pm 0.02$ \\
V3     & $1.0 {\pm} 0.2$  & $0.8 {\pm} 0.1$ & $1.2  {\pm} 0.2$  & $2.5 {\pm} 0.2$ & $3.2 {\pm} 0.2$ & $2.22 \pm 0.02$ \\
vdW    & $0.0$            & $0.5 {\pm} 0.1$ & $0.98 {\pm} 0.02$ & $2.7 {\pm} 0.2$ & $3.1 {\pm} 0.3$ & $2.33 \pm 0.02$ \\
\hline
\end{tabular}
\end{center}
\end{table*}

\begin{figure} [ht]
\centerline{\psfig{file=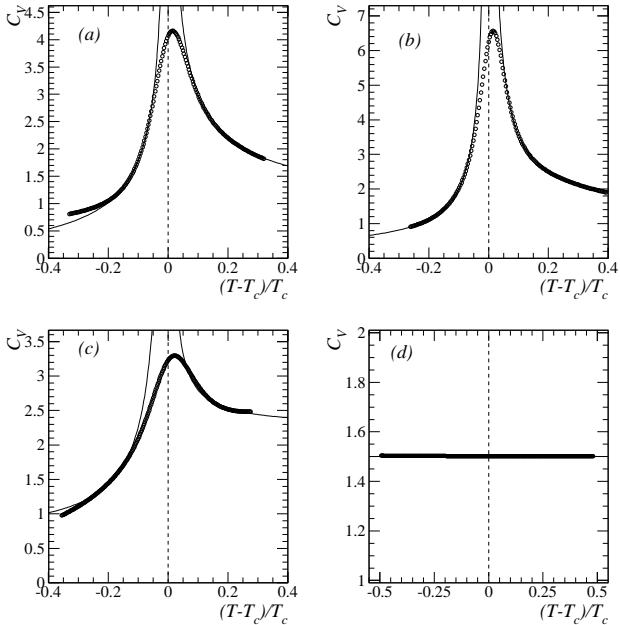,width=9.6cm,angle=0}}
\caption{Determination of the critical exponent $\alpha$ from the specific heat at a constant density, $\rho = \rho_c$ for (a)
V1, (b) V2, (c) V3 and (d) the van der Waals fluid.\ \ Solid lines show a sample fit.}
\label{fig:19}
\end{figure}

The functional form in eq. (\ref{alpha}) did not fit the curves shown in Figure 21a, b and c over the entire range of 
$( T - T_c ) / T_c$.\ \ To some degree this is to be expected.\ \ Near the critical point the finite size effects, which manifest 
themselves first in the smoothing of the kink of sub-critical free energy isochores, diminish a diverging specific heat into a 
peaking specific heat.\ \ Far from the critical point, the analytic terms in the expression for the specific heat become dominant 
and the power law behavior is overwhelmed.\ \ In some mid-range region, neither to0 far from nor too near to the critical point 
behavior consistent with eq. (\ref{alpha}) was observed.\ \ Various fits were tried on both sides of the critical point, but only 
those which gave a matching value for $\alpha$ were considered.\ \ Figures 21a, b and c show the results of one such fit.\ \ 
Table II lists the average results for many such fits.

The van der Waals fluid shows much different behavior that do the nuclear models.\ \ The constant value of $C_V = 3/2$ in the van
der Waals fluid leads to the result of $\alpha = 0$ as expected for a mean field model.\ \ Based on this result it would seem that
the nuclear models are not mean field models.\ \ They show a peaking in the specific heat that is inconsistent with the behavior of
a van der Waals fluid type of mean field model or the behavior of the Landau model which shows a discontinuity in the specific
heat.

The exponent $\beta$ is determined using the $( P, V_f )$ points along the coexistence curve, shown in Figure 15, which should be
described by
	\begin{equation}
		{\rho_{l}} - {\rho_{g}} \sim \left( \frac{T_c-T}{T_c} \right)^{\beta}
	\label{beta}
	\end{equation}
Fitting ${\rho}_l - {\rho}_g$ versus $\left( T - T_c \right) / T_c$ to a simple power law for the positive slope portion of 
Figure 22a, b and c gives the exponent ${\beta}$.\ \ See Table II for results.\ \ The van der Waals fluid recovers the mean field 
value of $\beta = 1/2$.\ \ The model V1 gives the least impressive fit results.

\begin{figure} [ht]
\centerline{\psfig{file=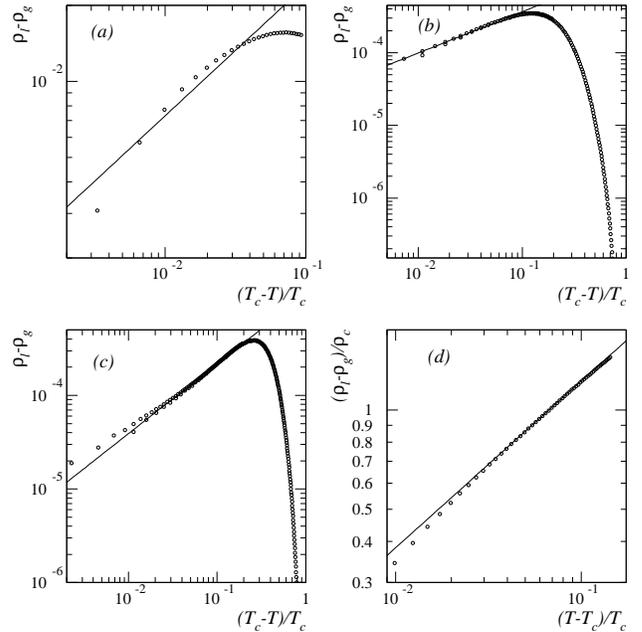,width=9.6cm,angle=0}}
\caption{Determination of the critical exponent $\beta$ from the liquid-gas density difference along the coexistence curve for (a)
V1, (b) V2, (c) V3 and (d) the van der Waals fluid.\ \ Solid lines show a sample fit.}
\label{fig:20}
\end{figure}

Near the critical point the isothermal compressibility, ${\kappa}_T$, is given by
	\begin{equation}
		{\kappa}_{T} = {\Gamma}_{\pm} \left| \frac{T-T_c}{T_c} \right|^{{\gamma}_{\pm}}
	\label{gamma}
	\end{equation}
For $T < T_c$ fitting ${\kappa}_T$ versus $\left| ( T - T_c ) / T_c \right|$ along the coexistence curve gives ${\gamma}_{-}$ while
${\gamma}_{+}$ is determined by fitting ${\kappa}_T$ versus $\left| ( T - T_c ) / T_c \right|$ for $T > T_c$ at ${\rho} =
{\rho}_c$.\ \ Due to the imprecise nature of the data along the coexistence curve for $T < T_c$ fits were made only for $T > 
T_c$.\ \ Fits were made over the entire region of $\left| ( T - T_c ) / T_c \right|$ of $T > T_c$.\ \ The results for the 
extraction of the exponent $\gamma$ are shown in Figure 23.

\begin{figure} [ht]
\centerline{\psfig{file=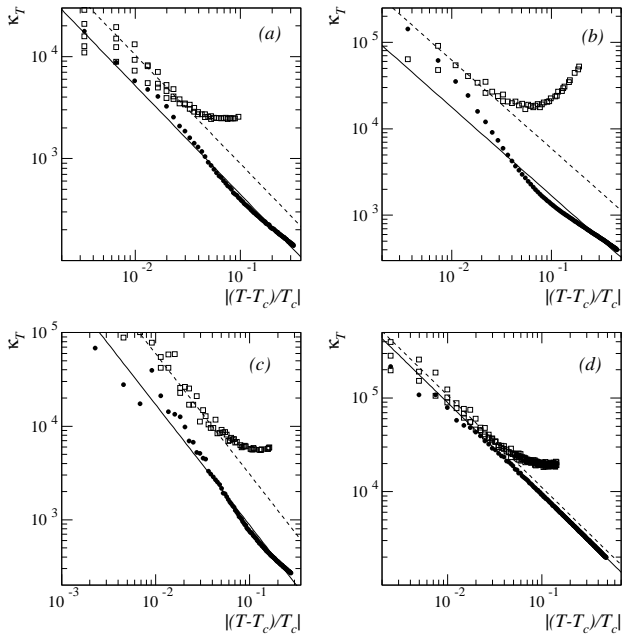,width=9.6cm,angle=0}}
\caption{Determination of the critical exponent $\gamma$ from the isothermal compressibility, ${\kappa}_T$, for (a) V1, (b) V2, (c)
V3 and (d) the van der Waals fluid.\ \ Solid circles show the behavior of ${\kappa}_T$ for $T > T_c$ and at $\rho = {\rho}_c$.\ \
Solid lines show a sample fit.\ \ Open squares show the behavior of ${\kappa}_T$ along one edge of the coexistence boundary.\ \
Dashed lines show the fit for $T > T_c$ multiplied by a constant.}
\label{fig:21}
\end{figure}

For V1 two different power law regions appear to be present, one close to the critical point and one further from the critical
point.\ \ See Figure 23a.\ \ The error bars on the $\gamma$-value in Table II account for this behavior.\ \ The fit to the entire
$\left| ( T - T_c ) / T_c \right|$ region is used because the resulting power law shows some agreement with the behavior of
${\kappa}_T$ for $T < T_c$ when the coefficient of the power law is increased by some factor.\ \ See dashed line and open squares
in Figure 23a.\ \ Similar arguments apply to the results for V2 and V3.\ \ See Figures 21b and c.

The van der Waals fluid shows the expected behavior and recovers the value of $\gamma = 1$ to within error bars.\ \ See Figure 23d
and Table II for results.\ \ The $T < T_c$ behavior of ${\kappa}_T$ also shows the expected power law behavior with the appropriate
exponent value.

Examining the shape of the critical isotherm leads to an estimation of the exponent $\delta$ from:
	\begin{equation}
		\left| P - P_c \right| \sim \left| \frac{{\rho}-{\rho}_c}{{\rho}_c} \right| ^{{\delta}_{\pm}} .
	\label{delta}
	\end{equation}
The critical isotherm was examined independently for ${\rho} < {\rho}_c$, which gives ${\delta}_{-}$, and ${\rho} > {\rho}_c$,
which gives ${\delta}_{+}$.\ \ As with the exponents $\alpha$ and $\gamma$, a system with a continuous phase transition the values
of ${\delta}_{\pm}$ should be the same on both sides of the critical density.\ \ This fact is again used as guide in searching for
fitting regions to extract the exponent $\delta$.\ \ See Figure 24 and Table II.

\begin{figure} [ht]
\centerline{\psfig{file=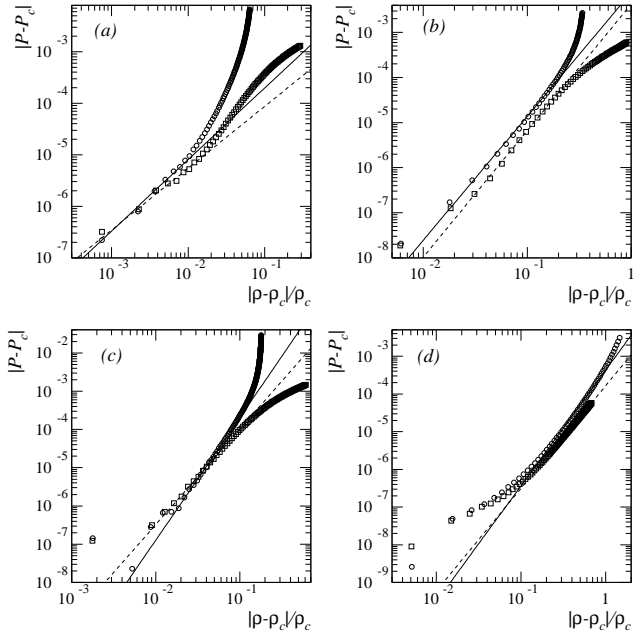,width=9.6cm,angle=0}}
\caption{Determination of the critical exponent $\delta$ from the critical isotherm for (a) V1, (b) V2, (c) V3 and (d) the van der
Waals fluid.\ \ Open circles show the critical isotherm for $\rho > \rho_c$ and at $\rho = \rho_c$.\ \ Solid lines show a
sample fit.\ \ Open squares show the critical isotherm for $\rho < \rho_c$.\ \ Dashed lines show a sample fit.}
\label{fig:22}
\end{figure}

For V1 only the regions closest to the critical point gave matching $\delta$-values.\ \ The error bars on the $\delta$-values in
Table II reflect the changes in ${\delta}_{\pm}$ when different fit regions are examined.\ \ In V2 there are regions on both sides 
of the critical point which yield a matching set of ${\delta}_{\pm}$ values.\ \ No such region could be found for V3, even very 
close to the critical point.\ \ The van der Waals fluid shows some regions on both sides of the critical point where 
${\delta}_{\pm}$ match, to within error bars, and agree with the expected value of $\delta = 3$.

Finally, the topological exponent, $\tau$, from Fisher's droplet model \cite{fisher} can be recovered based on considerations of 
the compressibility factor, $C_f$ via the relationship \cite{kiang}:
\begin{equation}
	C_f = \frac{{\zeta}({\tau})}{{\zeta}({\tau}-1)} .
\label{cf_tau}
\end{equation}
The Riemann $\zeta$ functions of eq. (\ref{cf_tau}) were summed from $1$ to $1 000 000 000$.\ \ When a value of $\tau = 7/3$ was input 
for the van der Waals fluid, eq. (\ref{cf_tau}) yielded a value of $0.393$ indicating that terminating the summation at $1 000 000 000$ 
yields a value of $C_f$ that is approximately $5$\% too high; for the van der Waals fluid $C_f = 3/8$.\ \ This supposition is supported 
by decreasing the upper summation limit and observing and increase in the value of $C_f$.\ \ This {\it error} was accounted for in the 
estimation of the value of $\tau$.\ \ See Table II for results.

\subsubsection{Scaling laws}

With four critical exponents determined it is possible to perform a consistency check using the well known scaling relations.\ \ 
For example, the Rushbrooke inequality shows that:
	\begin{equation}
		{\alpha} +2{\beta} + {\gamma} = 2 ,
	\label{scaling_1}
	\end{equation}
here shown as an equality in keeping with the scaling hypothesis and renormalization \cite{stanely_review}.\ \ And the Griffiths 
equality:
	\begin{equation}
		{\alpha} + {\beta}(1+{\delta}) = 2 
	\label{scaling_2}
	\end{equation}
and the Widom equality:
	\begin{equation}
		{\beta}({\delta}-1) - {\gamma} = 0 .
	\label{scaling_3}
	\end{equation}
And finally from Fisher's droplet model:
	\begin{equation}
		\frac{\beta}{\gamma} - \frac{\tau - 2}{3 - \tau} = 0 .
	\label{scaling_4}
	\end{equation}
Using the average values determined for ${\alpha}$, ${\beta}$, ${\delta}$, ${\gamma}$ and $\tau$ the results for these scaling laws 
are compiled in Table III.\ \ Only the van der Waals fluid results consistently satisfy the above scaling laws to within error bars.\ \
The nuclear models generally fail to satisfy three of the four scaling laws.\ \ This failure is inconsistent with the behavior of 
the phase diagram, shown for example in Figure 5, which appears show a critical point, thus indicating the presence of a continuous 
phase transition.\ \ While moderately good fits are observed for the specific heat, the {\it liquid}-{\it gas} density difference, the 
isothermal compressibility and the critical isotherm for each of the versions of the nuclear model, the meaning of these power 
laws and critical exponents remains an open question in light of the failure to adhere to well known scaling laws.

\begin{table*}
\begin{center}
\caption{Scaling law results}
\begin{tabular}{lllll}
\hline 
System & Rushbrooke      & Griffiths       & Widom            & Fisher         \\
\hline                                                            
V1     & $2.6 {\pm} 0.4$ & $1.8 {\pm} 0.5$ & $ 0.9 {\pm} 0.2$ & $-0.5 \pm 0.2$ \\
V2     & $3.2 {\pm} 0.5$ & $3.2 {\pm} 0.4$ & $-0.1 {\pm} 0.2$ & $ 0.5 \pm 0.3$ \\
V3     & $3.7 {\pm} 0.3$ & $3.9 {\pm} 0.6$ & $-0.2 {\pm} 0.3$ & $ 0.4 \pm 0.1$ \\
vdW    & $2.0 {\pm} 0.1$ & $2.0 {\pm} 0.4$ & $-0.0 {\pm} 0.2$ & $ 0.0 \pm 0.1$ \\
\hline
\end{tabular}
\end{center}
\end{table*}

\section{Summary}

It has been shown that the type of nuclear model discussed here exhibits many features commonly associated with a system in which 
critical phenomena are present, {\it e.g.} a coexistence curve, power laws, critical exponents.\ \ By removing both the Coulomb and 
temperature-dependent free energy terms, it was found that the appearance of a critical point in these models is due to the interplay 
between the surface, volume, and translational free energy terms.\ \ However, these types of models are not without inconsistencies.\ \ 
One striking inconsistency is the fact that the temperature dependent surface free energy gives rise to a infinite negative specific heat 
at the critical temperature used by the model.\ \ Furthermore, no version of the model showed a critical temperature that agreed with the 
one explicitly input into the surface free energy term.\ \ Additionally, when the surface term was rendered temperature independent the 
critical point remained.\ \ Thus suggesting that the appearance of a critical point in these models is not dependent on the temperature 
dependence of the surface term but rather is a result of the interplay between the surface and volume free energy terms.

The critical temperature and density have been determined by examining isotherms in the $P - \rho$ plane.\ \ In the neighborhood of 
this critical point, singular behavior characterized by power laws was observed.\ \ However, these critical exponents do not obey well 
known scaling relations.\ \ This is a particularly troublesome occurrence as any model with true critical behavior, even the simple 
van der Waals fluid, does have exponents which obey these scaling relations.\ \ It is possible that an examination of this model for 
larger systems, with smaller steps in temperature and volume in the calculation of the free energy, will yield a consistent set of 
critical exponents.

It is important to note that the critical densities found here are much higher than could be realized with a closest packing of 
normal density nuclei.\ \ Additionally, these critical densities are significantly higher than those typically used to compare model 
predictions \cite{bondorf}, \cite{gross} to data.

A major conclusion of this work is that the particular phenomenological description of the free energy of a hot nucleus leads to several
inconsistencies regarding both temperature and density.\ \ It was pointed out that the parameterization of the surface free energy leads to a
negative and divergent contribution to the specific heat at $T$ approaches the value of the parameter $T_c$ in eq. (\ref{F_AZ_S}).\ \ 
Furthermore, all values of the critical temperature found from examination of isotherms in the $P$-$\rho$ plane are much below this parameter 
value.\ \ Thus, while use of such a model may well lead to an excellent description of multifragmentation data, the lack of internal consistency 
noted here makes the interpretation of data in terms of the model problematic.\ \ Such agreement may rest more on the phase space sampling and 
variable free volume inherent in the model than on the finer details examined here.

Finally, it has been shown that the variable volume version of this phenomenological model of multifragmentation exhibits caloric 
curves which can be misinterpreted in the absence of detailed knowledge of the complete thermodynamic phase diagram.

This work was supported in part by the U.S. Department of Energy Contracts or Grants No. DE-ACO3-76F00098, DE-FG02-89ER-40513,
DE-FG02-88ER-40408, DE-FG02-88ER40412, DE-FG05-ER40437 and by the U.S. National Science Foundation under Grant No. PHY-91-23301.

\end{document}